\documentclass[iop,apjl]{emulateapj}
\usepackage{ulem}

\newcommand{\simless}{\mathbin{\lower 3pt\hbox
      {$\rlap{\raise 5pt\hbox{$\char'074$}}\mathchar"7218$}}} 
\newcommand{\simgreat}{\mathbin{\lower 3pt\hbox
     {$\rlap{\raise 5pt\hbox{$\char'076$}}\mathchar"7218$}}} 

\slugcomment{\today}

\shorttitle{Brown dwarf disks with ALMA}
\shortauthors{Ricci et al.}

\begin{document}


\title{Brown dwarf disks with ALMA}


\author{L. Ricci\altaffilmark{1}, 
L. Testi\altaffilmark{2,3}, 
A. Natta\altaffilmark{3,4},
A. Scholz\altaffilmark{4},
I. de Gregorio-Monsalvo\altaffilmark{2,5} and
A. Isella\altaffilmark{1}}

\altaffiltext{1}{Department of Astronomy, California Institute of Technology, MC 249-17, Pasadena, CA 91125, USA}
\altaffiltext{2}{European Southern Observatory, Karl-Schwarzschild-Strasse 2, D-85748 Garching, Germany}
\altaffiltext{3}{INAF-Osservatorio Astrofisico di Arcetri, Largo E. Fermi 5, I-50125 Firenze, Italy}
\altaffiltext{4}{School of Cosmic Physics, Dublin Institute for Advanced Studies, 31 Fitzwilliam Place, Dublin 2, Ireland}
\altaffiltext{5}{Joint ALMA Observatory (JAO)/ESO. Alonso de Cordova 3107. Vitacura 763 0335. Santiago de Chile}
\email{lricci@astro.caltech.edu}

\begin{abstract}

\noindent We present ALMA continuum and spectral line data at 0.89~mm and 3.2~mm for three disks surrounding young brown dwarfs and very low mass stars in the Taurus star forming region. Dust thermal emission is detected and spatially resolved for all the three disks, while CO($J=3-2$) emission is seen in two disks. We analyze the continuum visibilities and constrain the disks physical structure in dust. The results of our analysis show that the disks are relatively large, the smallest one with an outer radius of about 70~AU. The inferred disk radii, radial profiles of the dust surface density and disk to central object mass ratios lie within the ranges found for disks around more massive young stars. 
We derive from our observations the wavelength dependence of the millimeter dust opacity.
In all the three disks data are consistent with the presence of grains with at least millimeter sizes, as also found for disks around young stars, and confirm that the early stages of the solid growth toward planetesimals occur also around very low mass objects.
We discuss the implications of our findings on models of solids evolution in protoplanetary disks, on the main mechanisms proposed for the formation of brown dwarfs and very low mass stars, as well as on the potential of finding rocky and giant planets around very low mass objects.


\end{abstract}


\section{Introduction}
\label{sec:intro}

\noindent For several million years, young stellar objects are surrounded by dusty disks which are the remnants of the cloud cores from which stars form and at the same time the mass reservoir for the formation of planets. Similarly, young brown dwarfs -- objects intermediate in mass between stars and planets -- harbor dusty disks \citep{Comeron:1998, Natta:2001, Muench:2001, Natta:2002}. Over the past decade, more than one hundred brown dwarf disks have been detected and analysed in the infrared \citep[e.g.][]{Jayawardhana:2003, Liu:2003, Scholz:2007, Harvey:2012}. These ``circum-sub-stellar'' disks show the same signs of evolution than the disks around young stellar objects -- downsloping SEDs typical for flat disks \citep[e.g.][]{Scholz:2007}, silicate features indicating grain growth in the inner disk \citep[e.g.][]{Sterzik:2004, Apai:2005}, and inner disk clearing \citep[e.g.][]{Muzerolle:2006}. 

Most of these results, however, relate only to a small fraction of the dust in the inner parts of the disks. Investigating the bulk of the dust, which is expected to reside in the cold outer regions of the disk, is considerably more difficult and requires observations at longer sub-mm/mm wavelengths. 
Since at these wavelengths disk continuum fluxes roughly scale with object mass \citep{Scholz:2006, Mohanty:2013, Andrews:2013}, substellar disks are very faint. So far, only a handful of bright brown dwarf disks are detected in this wavelength domain. For that reason, observational constraints on the global disk properties are sparse. This is especially true for the sizes of substellar disks -- only one direct measurement \citep{Ricci:2013} and some more indirect constraints \citep{Scholz:2006, Luhman:2007} are available. The limited information points to relatively small disks with maximum radii of 10-50\,AU and masses up to $\sim$ a few $M_{\mathrm{Jup}}$.

There are strong incentives for sub-mm/mm observations of brown dwarf disks. On one side, the size and mass of disks may carry information about the early evolutionary stages of brown dwarfs. In hydrodynamical simulations, the formation of brown dwarfs is linked with dynamical interactions in stellar clusters which eject the nascent brown dwarfs from its accretion reservoir and thus prevent it from becoming a star \citep{Reipurth:2001, Bate:2003}. Alternatively, brown dwarfs may get ejected from disks around stars, either as embryos \citep{Basu:2012} or as fully formed objects \citep{Stamatellos:2009}. If close encounters with other objects are essential for brown dwarf formation, one could expect their disks to be truncated or depleted. For example, early findings by \citet{Bate:2003} indicate that the overwhelming majority of brown dwarfs is expected to have only very small disks ($<20$\,AU). In the more recent simulations by \citet{Bate:2009} and \citet{Bate:2012}, most of the disks around very low mass objects have truncation radii smaller than 40\,AU. This should still be treated as a lower limit, because the simulations produce a dense stellar cluster with multiple interactions per object and they do not follow the disk evolution to ages of several Myr. 

On the other side, disks are the sites of planet formation. The properties of brown dwarf disks provide useful information on planet formation in extreme environments and possibly on the ubiquity of planetary systems. In all scenarios for planet formation, the total mass of the disk is a crucial limit for the mass of the planets that can form in such a disk. Earth-mass planets are expected to form readily by core accretion in brown dwarf disks of a few Jupiter masses. If the disks are significantly less massive, the masses of the resulting planets would scale down as well \citep{Payne:2007}. Current planet surveys are beginning to probe the regime of very low mass stars and indicate that Earths and super-Earths are common around M dwarfs \citep{Dressing:2013, Bonfils:2013}. Systematic searches for planets orbiting brown dwarfs are within our grasp \citep{Belu:2013}, which makes the exploration of their potential birth environment even more compelling.

The new Atacama Large Millimeter/submillimeter Array (ALMA) will provide important observational benchmarks for brown dwarf disks \citep{Natta:2008}. Even with the limited array available in Cycle 0, ALMA in Early Science was able to easily detect and resolve brown dwarf disks. In our first ALMA paper based on Cycle 0 data we presented strong evidence for grain growth in a disk around the brown dwarf $\rho$\,Oph ISO-102 \citep{Ricci:2012a}. Here we present new ALMA observations for three young brown dwarfs in the Taurus star forming region.

\noindent 
\section{ALMA Observations}
\label{sec:obs}

\noindent We observed 2M0444 (2MASS J04442713+2512164), CIDA 1 (2MASS J04141760+2806096) and CFHT Tau 4 (2MASS J04394748+2601407) using ALMA Early Science in Cycle~0 at Band~7 and~3 (about 338 and 93~GHz, respectively). 
Some of the main information on the observations are presented in Table~\ref{tab:obslog}.

\begin{table*}
\centering \caption{Summary of the ALMA observations. } \vskip 0.1cm
\begin{tabular}{ccccccc}
\hline
\hline

\\ 
Date  &     Target sources      &    ALMA Band    & Flux Calibrator & Antennas & PWV & baseline lengths \vspace{1mm} \\
   &                                      &                             &                             &                   &      (mm)   & (m)          \\
  
\\
\hline
\\
2012 Aug 27 & CIDA 1, 2M0444, CFHT Tau 4 &        7        &     Ganymede      &    25 & 0.60-0.85 & 21.7-402.3  \vspace{1mm} \\
2012 Nov 18  & CIDA 1, 2M0444                         &        3        &    Callisto             &    25 & 0.65-0.75 & 21.2-374.7   \vspace{1mm} \\
2012 Nov 19  & CFHT Tau 4                                 &        3        &     Ganymede      &    26 & 0.40-0.55 & 15.1-374.7  \vspace{1mm} \\
2012 Nov 20  & CFHT Tau 4                                 &        3        &     Ganymede      &    25 & 0.45-0.50 & 15.1-374.7  \vspace{1mm} \\

\\
\hline
\end{tabular}

\label{tab:obslog}

\end{table*}

All observations were performed with either 25 or 26 antennas in the Extended array configuration, which provides projected baseline lengths up to $\sim$400~m.
All observations were done in good and stable weather conditions, with precipitable water vapor always below $\sim 0.9$ mm (see Table~\ref{tab:obslog}).
The ALMA correlator was set to record dual polarization with four separate spectral windows, each providing an effective bandwidth of 1.875~GHz with channels of 0.488~MHz width. Spectral windows were centered at 331.102, 332.998, 343.102, 344.998 GHz for Band~7 observations and 86.102, 87.998, 98.103, 99.998 GHz for Band~3.
The total integration times on the three disks were approximately: 9 minutes for CIDA 1 and CFHT Tau 4 and 5 minutes for 2M0444 in Band 7;  15 minutes for CIDA 1 and 2M0444 and 55 minutes for CFHT Tau 4 in Band~3.

The interferometric visibility data were calibrated with the 3.4.0 version of the CASA software package \citep[][]{McMullin:2007}. The spectral response of the system was corrected using observations of the bandpass calibrator J0538-440, whereas Ganymede and Callisto were observed for flux calibration. Simultaneous observations of the 183~GHz water line with the water vapor radiometers were used to reduce atmospheric phase noise before using J0510+180 for standard complex gains calibration. The antenna-based phase noise over the bandpass scan showed median values of 7.1 and 2.7 degrees before and after the water vapor radiometer correction, respectively.
Self-calibration could be performed only in the case of CIDA 1, which is the brightest among our sources. The flux scale was tied to the Butler-JPL-Horizons 2012 models of Ganymede and Callisto, resulting in an accuracy of  $\sim10\%$. 

Imaging of the calibrated visibilities was done in CASA. For each band, a continuum map was produced using natural weighting and combining all the channels without line emission. The only line emission detected was the CO ($J=3-2$) rotational line in Band~7 for 2M0444 and CIDA 1. 

\section{Results of the observations}
\label{sec:res}

\subsection{Continuum maps and visibilities}
\label{sec:cont}

\begin{figure*}
\centering
\begin{tabular}{cc}
\includegraphics[scale=0.77,trim=100 0 0 0 0]{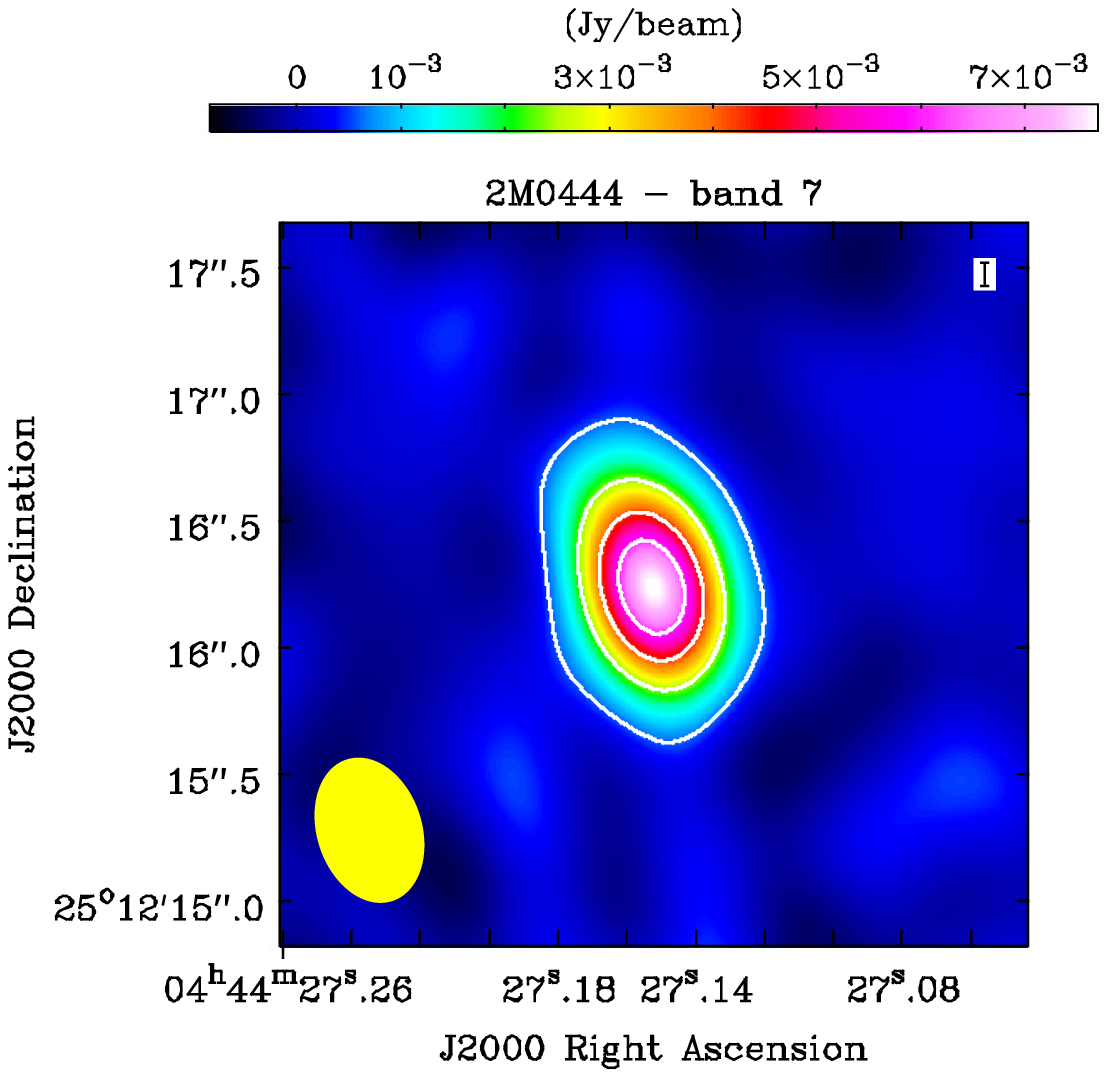} & 
\includegraphics[scale=0.5,trim=200 0 0 0]{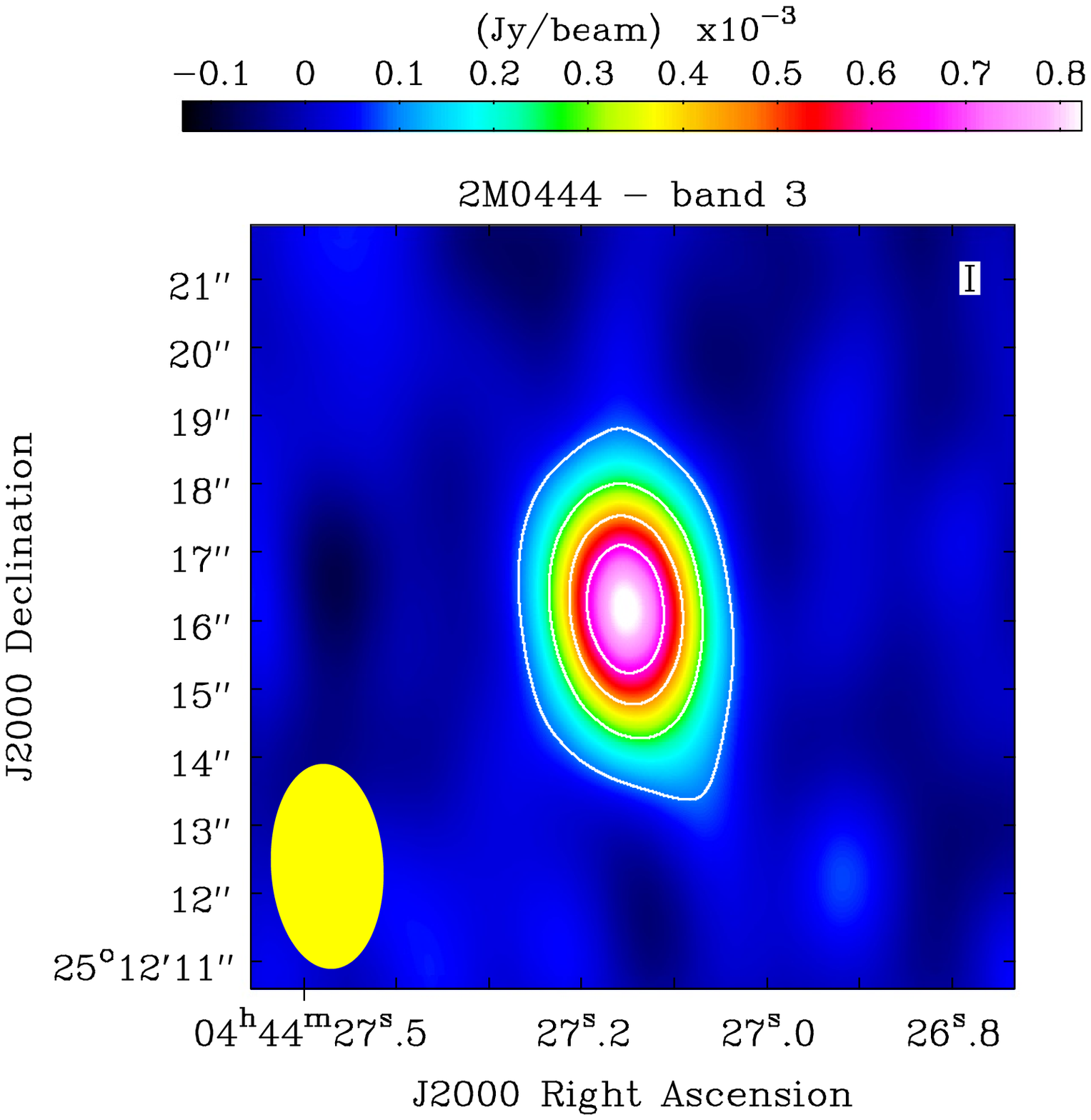}
\end{tabular}
\caption{ALMA continuum maps of 2M0444 in Band 7 (left panel) and Band 3 (right). In the left panel contours are drawn at 3, 12, 21, 30$\sigma$, with $1\sigma = 0.2$~mJy/beam. Synthesized beam is shown in the lower left corner and has sizes of $0.58'' \times 0.41''$ with P.A. $= 17.7$ degrees. In the right panel contours are at 3, 9, 15, 21$\sigma$, with $1\sigma = 0.03$~mJy/beam. Beam sizes are $2.99'' \times 1.63''$ with P.A. $= 3$ degrees.} 
\label{fig:cont_map_2m0444}
\end{figure*}

\begin{figure*}
\centering
\begin{tabular}{cc}
\includegraphics[scale=0.52,trim=50 0 0 0]{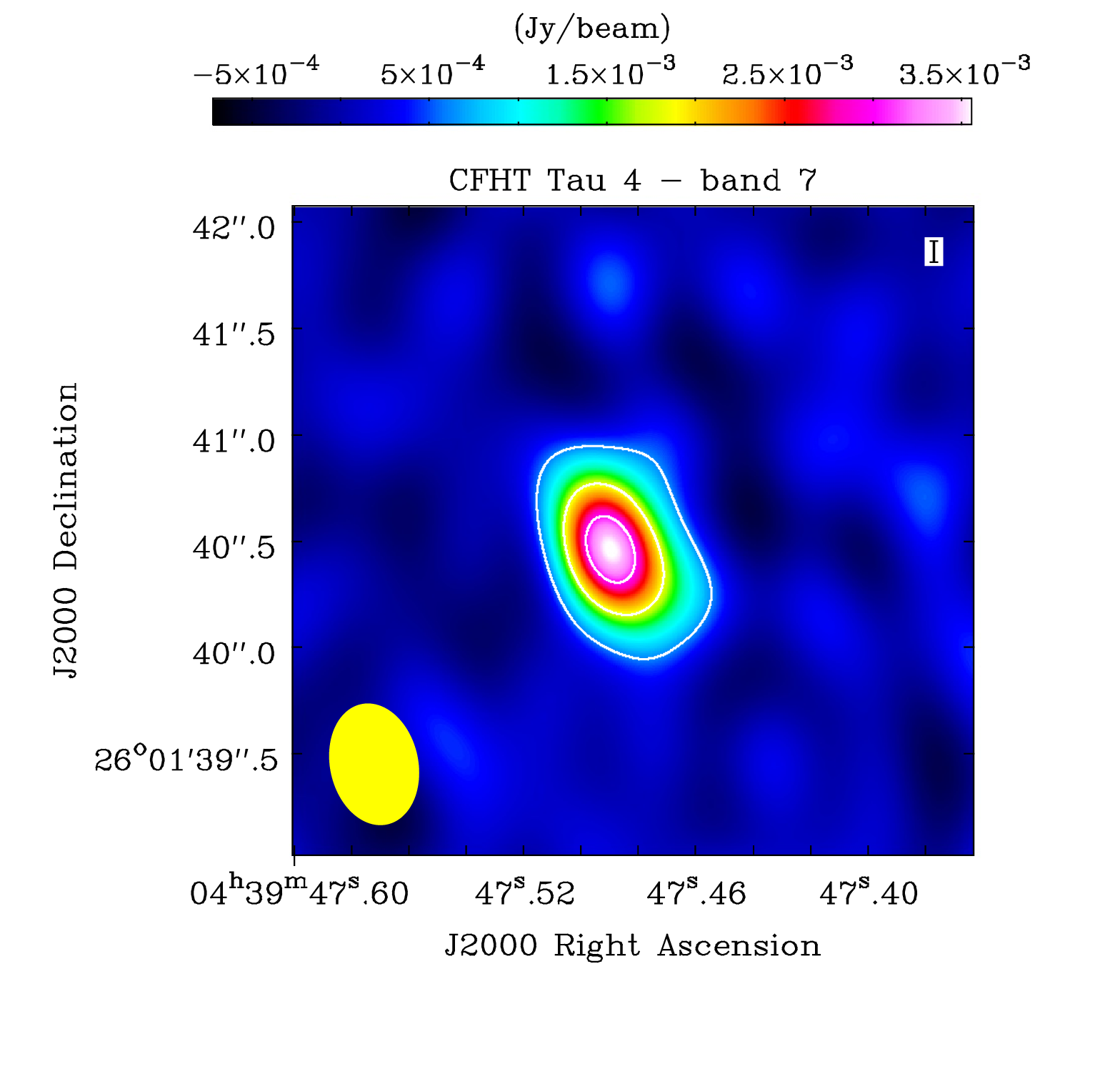}  &
\includegraphics[scale=0.5,trim=50 0 0 0]{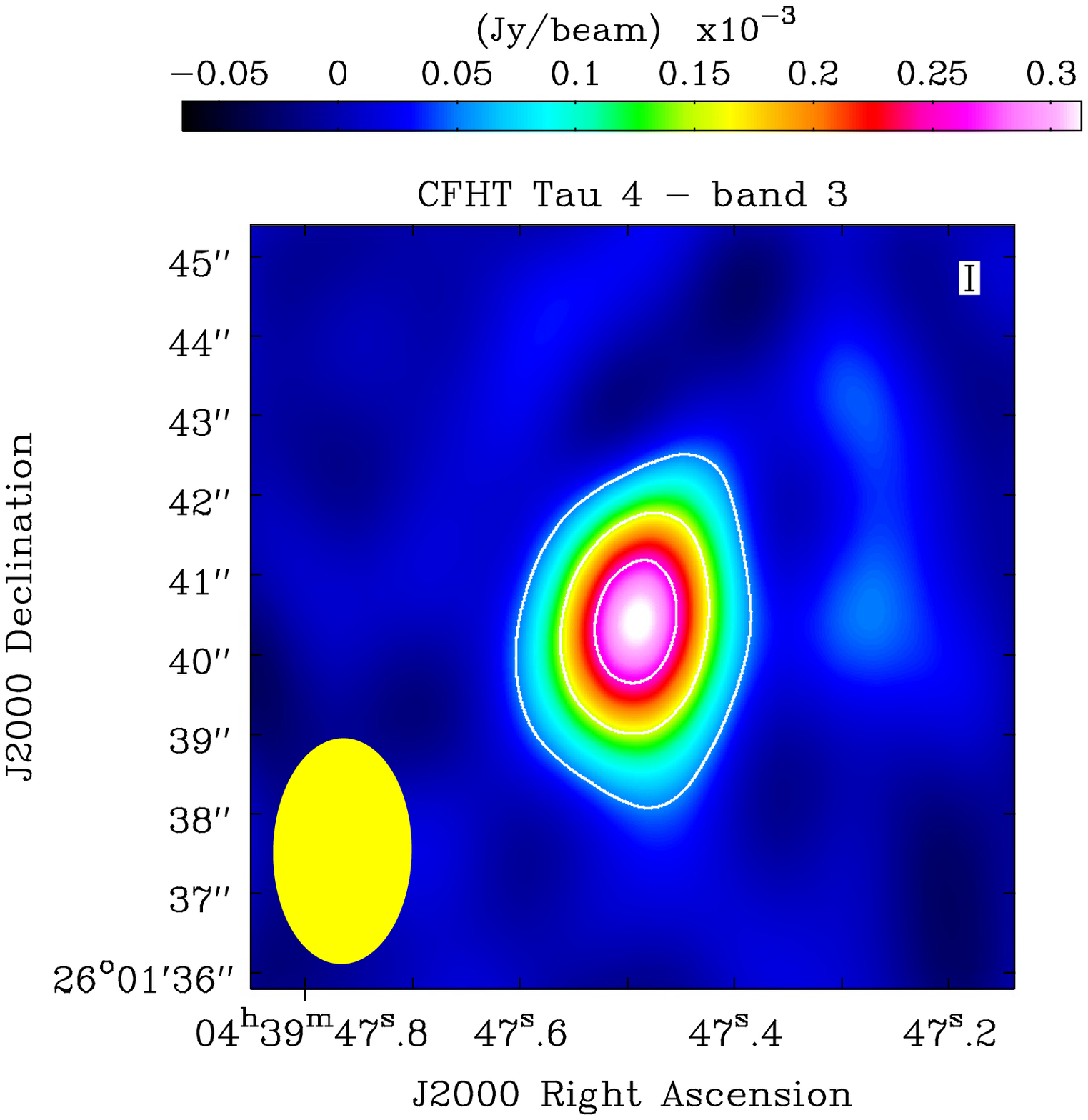} 
\end{tabular}
\caption{ALMA continuum maps of CFHT Tau 4 in Band 7 (left panel) and Band 3 (right).  In the left panel contours are drawn at 3, 9, 15, 21$\sigma$, with $1\sigma =$ 0.2 mJy/beam. Synthesized beam is shown in the lower left corner and has sizes of $0.57'' \times 0.41''$ with P.A. $= 12.2$ degrees.  In the right panel contours are at  3, 9, 15$\sigma$, with $1\sigma =$ 0.017 mJy/beam. Beam sizes are $2.82'' \times 1.73''$ with P.A. $= 179$ degrees.}
\label{fig:cont_map_cfht4}
\end{figure*}

\begin{figure*}
\centering
\begin{tabular}{cc}
\includegraphics[scale=0.52,trim=50 0 0 0]{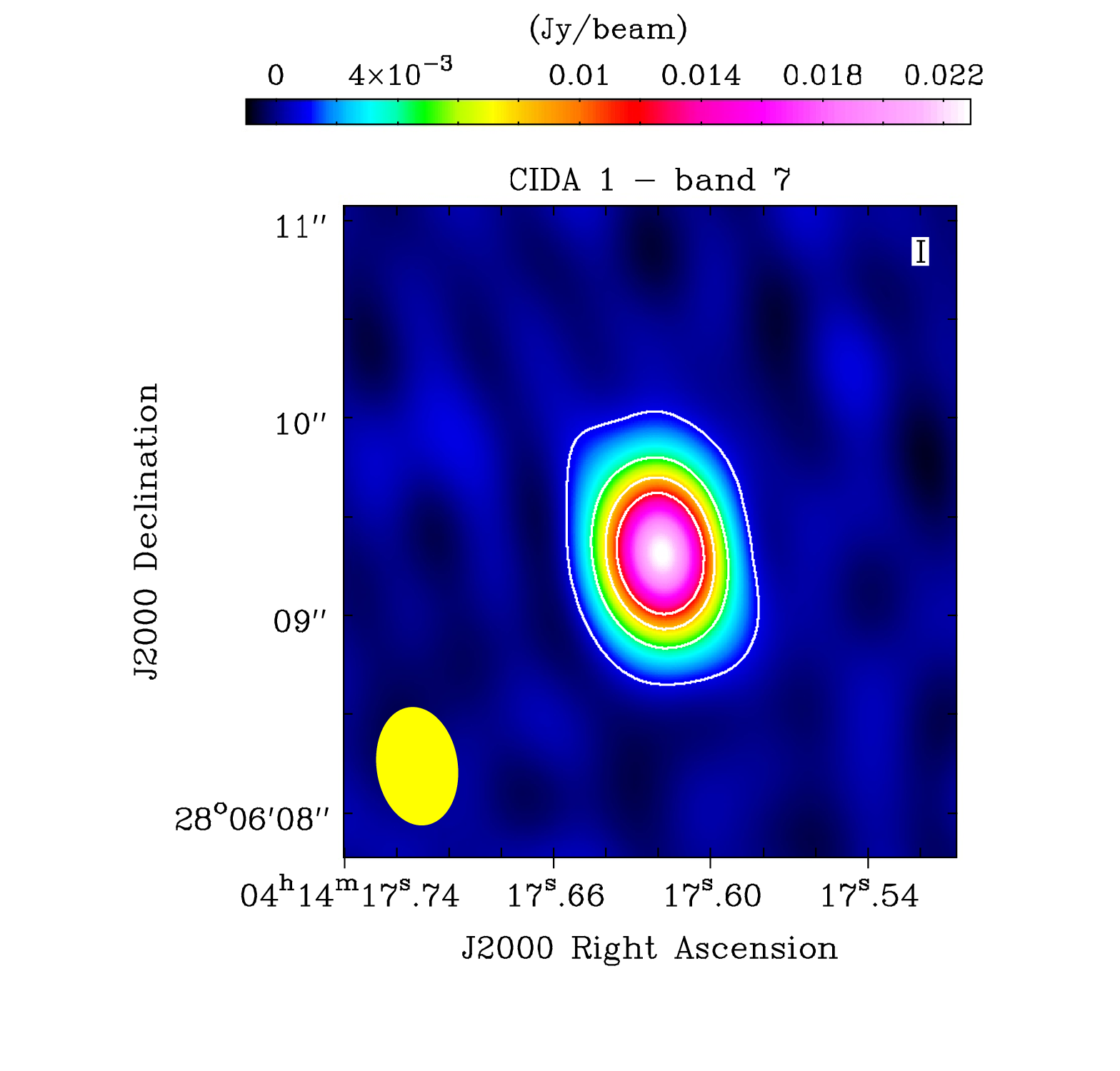} &
\includegraphics[scale=0.5,trim=50 0 0 0]{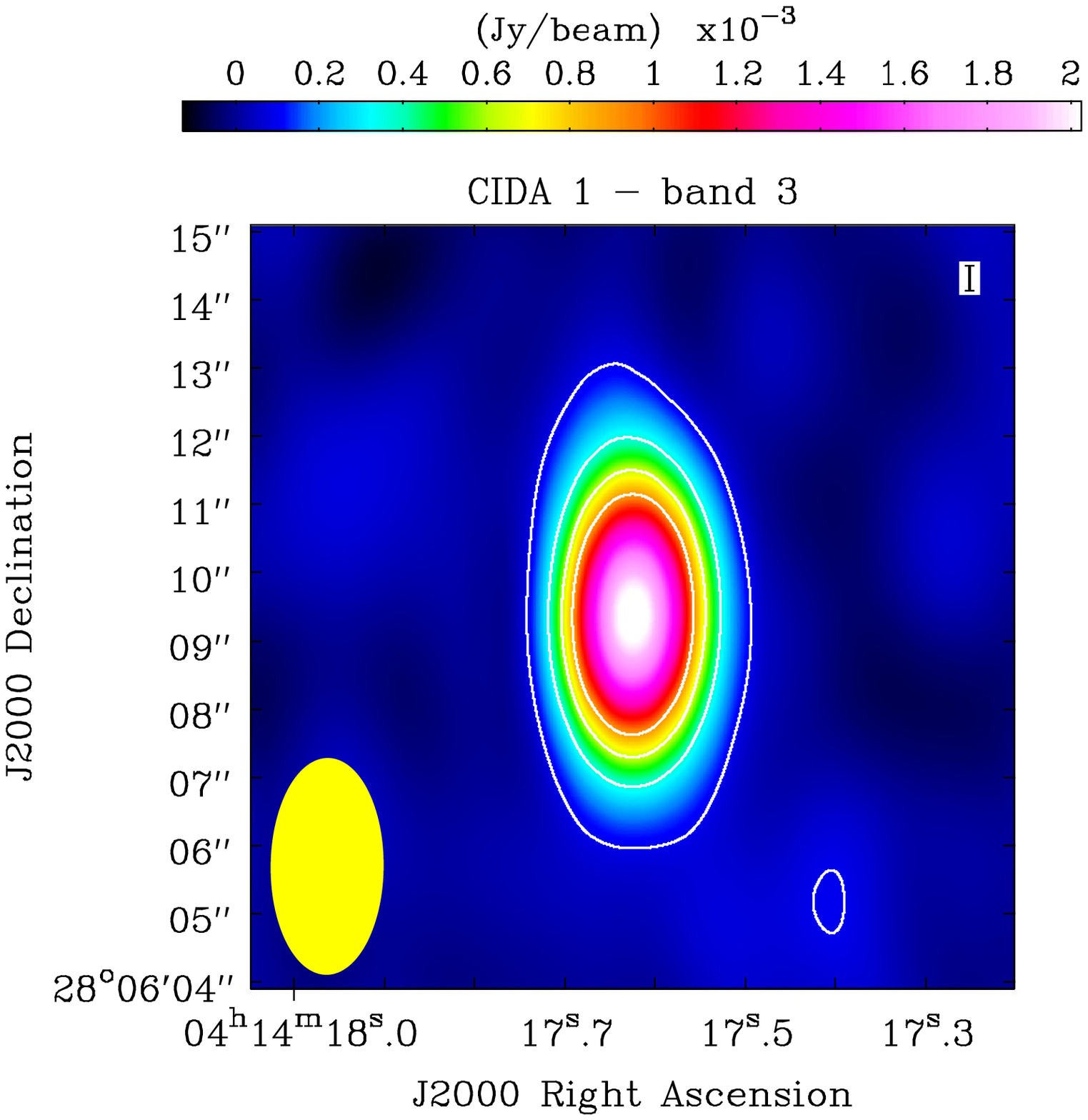} 
\end{tabular}
\caption{ALMA continuum maps of CIDA 1 in Band 7 (left panel) and Band 3 (right).  In the left panel contours are drawn at  3, 15, 27, 39$\sigma$, with $1\sigma =$ 0.3 mJy/beam. Synthesized beam is shown in the lower left corner and has sizes of $0.60'' \times 0.40''$ with P.A. $= 8.2$ degrees.  In the right panel contours are at  3, 12, 21, 30$\sigma$, with $1\sigma =$ 0.03 mJy/beam.  Beam sizes are  $3.16'' \times 1.64''$ with P.A. $= 0$ degrees.}
\label{fig:cont_map_cida1}
\end{figure*}

\noindent The maps of the dust continuum emission in Band 7 (0.89~mm) and 3 (3.2~mm) for 2M0444, CFHT Tau 4 and CIDA 1 are shown in Figures~\ref{fig:cont_map_2m0444}-\ref{fig:cont_map_cida1}.
The dust emission was detected for all the three disks in both bands. The measured flux densities at the center frequency of each ALMA Band obtained by integrating the surface brightness from each pixel with emission, and correspondent rms noise in each map are reported in Table~\ref{tab:fluxes}. The signal-to-noise ratio of these observations range from about 20 for CFHT Tau 4 to about 90 for CIDA 1. 
From these fluxes, an estimate of the spectral index of the SED between these frequencies can be derived through $\alpha_{\rm{338-93 GHz}} = \frac{log(F_{\rm338GHz}/F_{\rm93GHz})}{log({\rm338GHz}/{\rm93GHz})}$. The derived values are listed in the last column of Table~\ref{tab:fluxes}, where the reported uncertainties account for both the noise level in the maps and the $10\%$ uncertainty on the determination of the absolute flux scale (see Section~\ref{sec:obs}).  

\begin{figure*}[t!]
\centering
\begin{tabular}{cc}
\includegraphics[scale=0.5,trim=50 0 0 0]{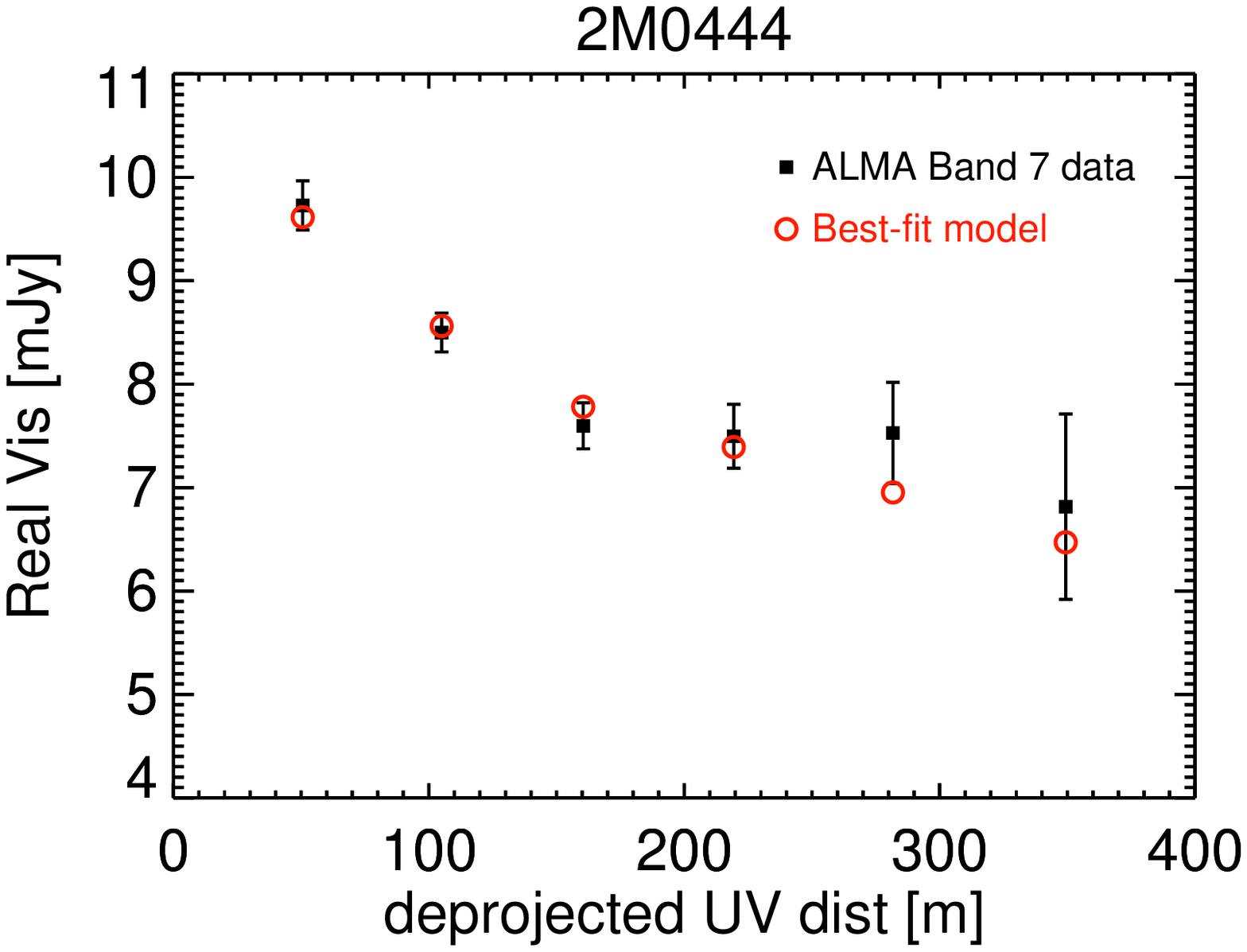} &
\includegraphics[scale=0.5,trim= 0 0 0 0]{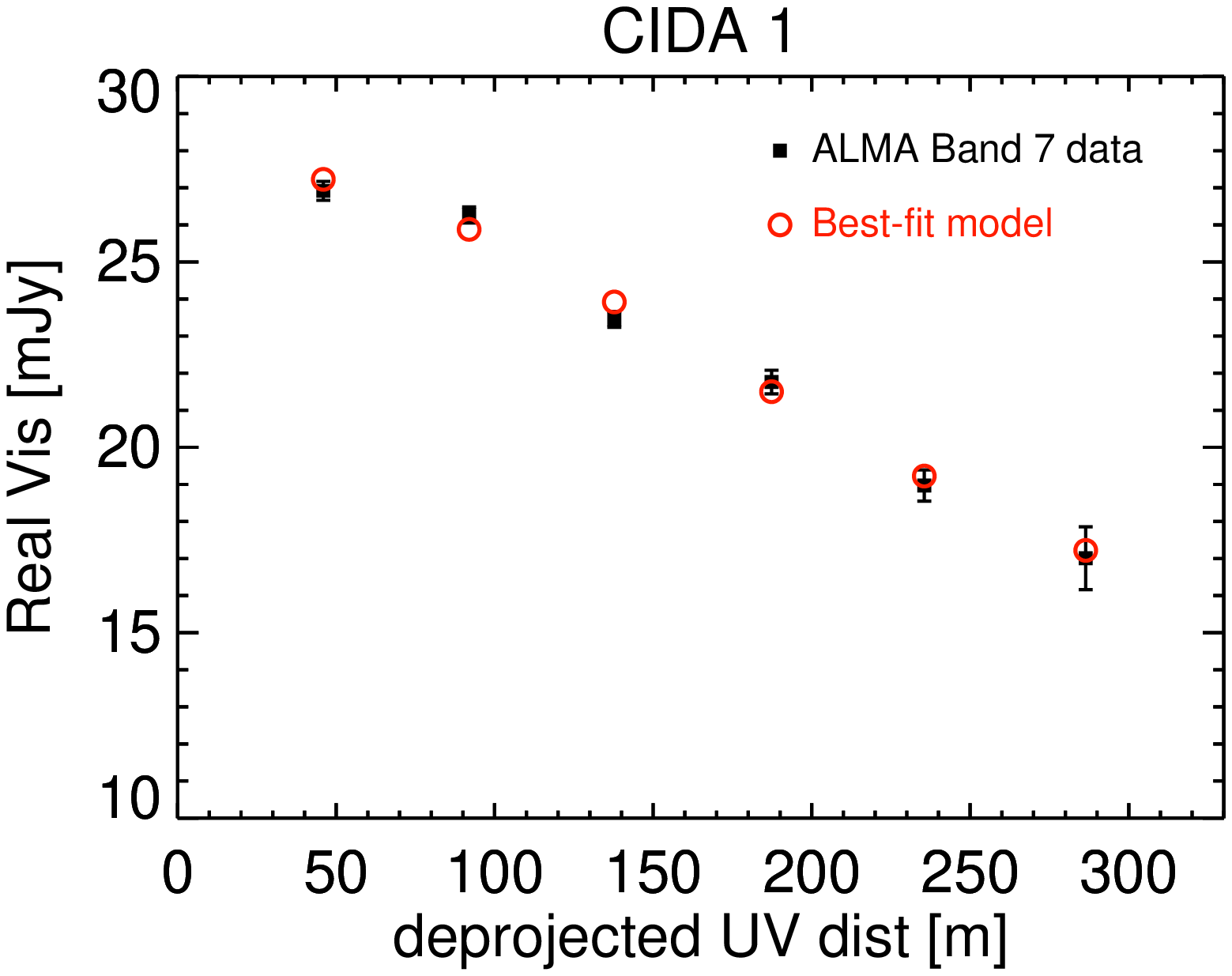} \\
 \multicolumn{2}{c}{\includegraphics[scale=0.5,trim= 0 0 0 0]{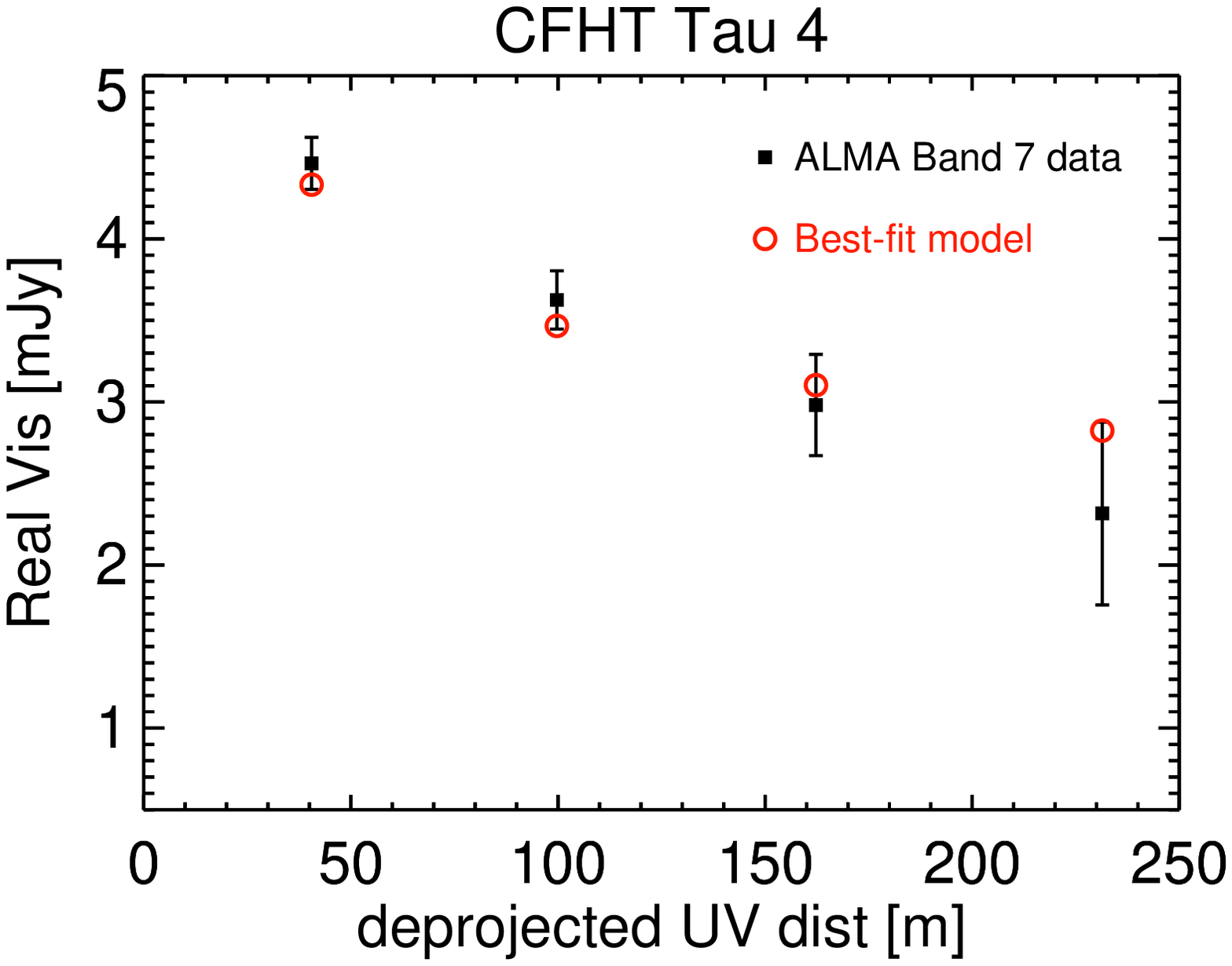}}
\end{tabular}
\caption{Real part of the visibility function vs deprojected baseline length for the ALMA observations of the three disks in Band 7. Each black data point shows the weighted mean, and uncertainty, after binning over deprojected baseline lengths. Red symbols show the real part of the visibility function for the best-fit disk models as described in Section~\ref{sec:visibilities}.}
\label{fig:amp_vis}
\end{figure*}

Figure~\ref{fig:amp_vis} shows the real part of the interferometric visibility function as measured by our ALMA observations in Band 7 plotted against the deprojected baseline lengths. For all the 3 disks the visibility amplitudes show a clear decrease at longer baseline lengths, indicating that the dust emission is spatially resolved. Given the angular size of the shortest scales probed by our observations, i.e. about $\approx$ 0.40$''$, this implies that the 3 disks have physical sizes in dust larger than about 56 AU in diameter or 28 AU in radius at the Taurus distance of 140~pc.  
More quantitative constraints on the physical structure of the disks from these observations are derived in Section~\ref{sec:visibilities}.

The visibility amplitudes in Band 3 are roughly constant with projected baseline length and are not shown here. As the projected baseline lengths used in those observations are very similar to those for the Band 7 observations (see Table \ref{tab:obslog}), the angular resolution is poorer by a factor of $\approx 338/93 = 3.6$ because of the lower frequency, and not high enough to spatially resolve the disk at those wavelengths.

\subsection{CO maps}
\label{sec:co}

\begin{figure*}
\centering
\begin{tabular}{cc}
\includegraphics[scale=0.5,trim=60 0 0 0]{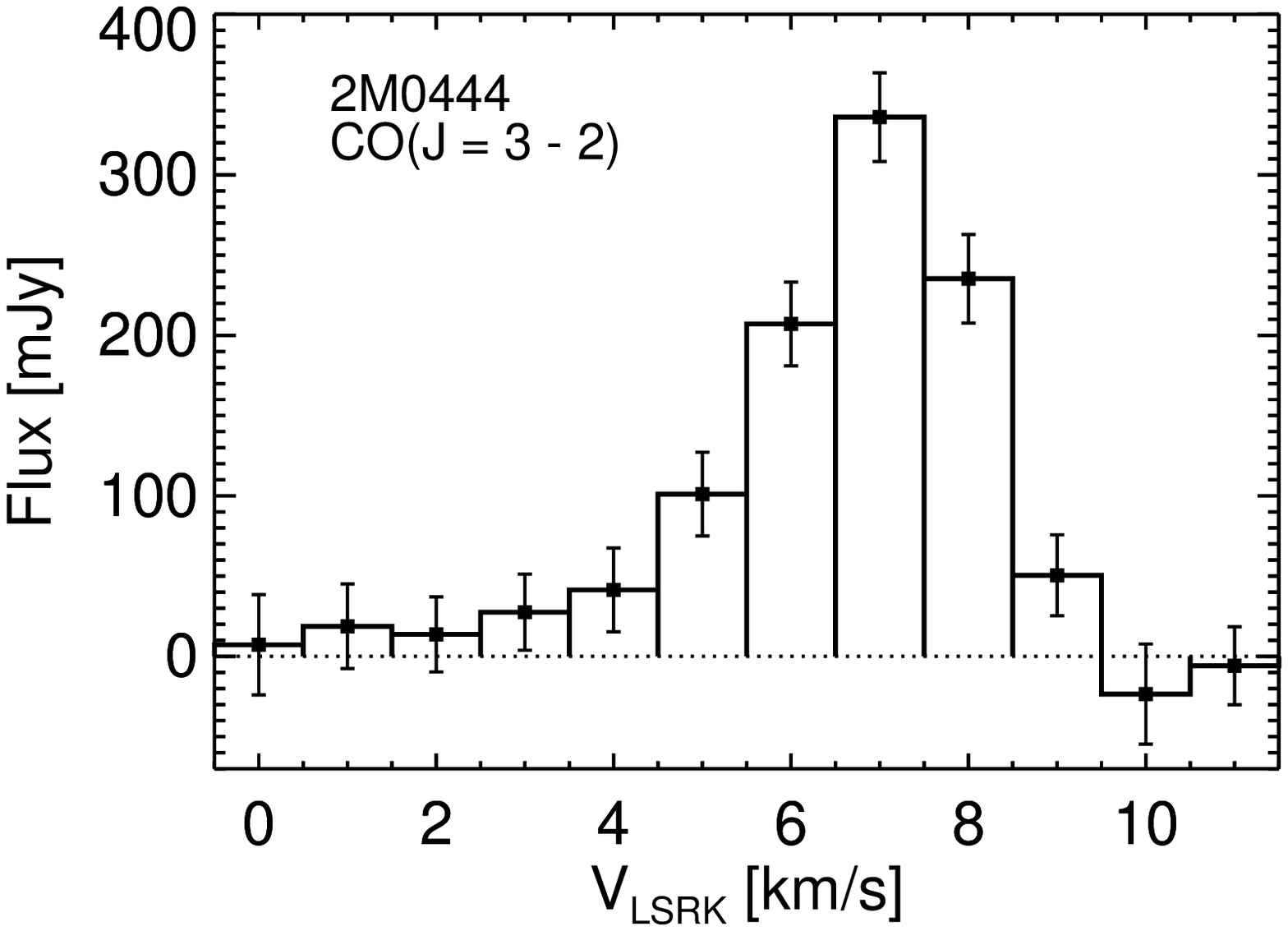} & 
\includegraphics[scale=0.5,trim=0 0 0 0]{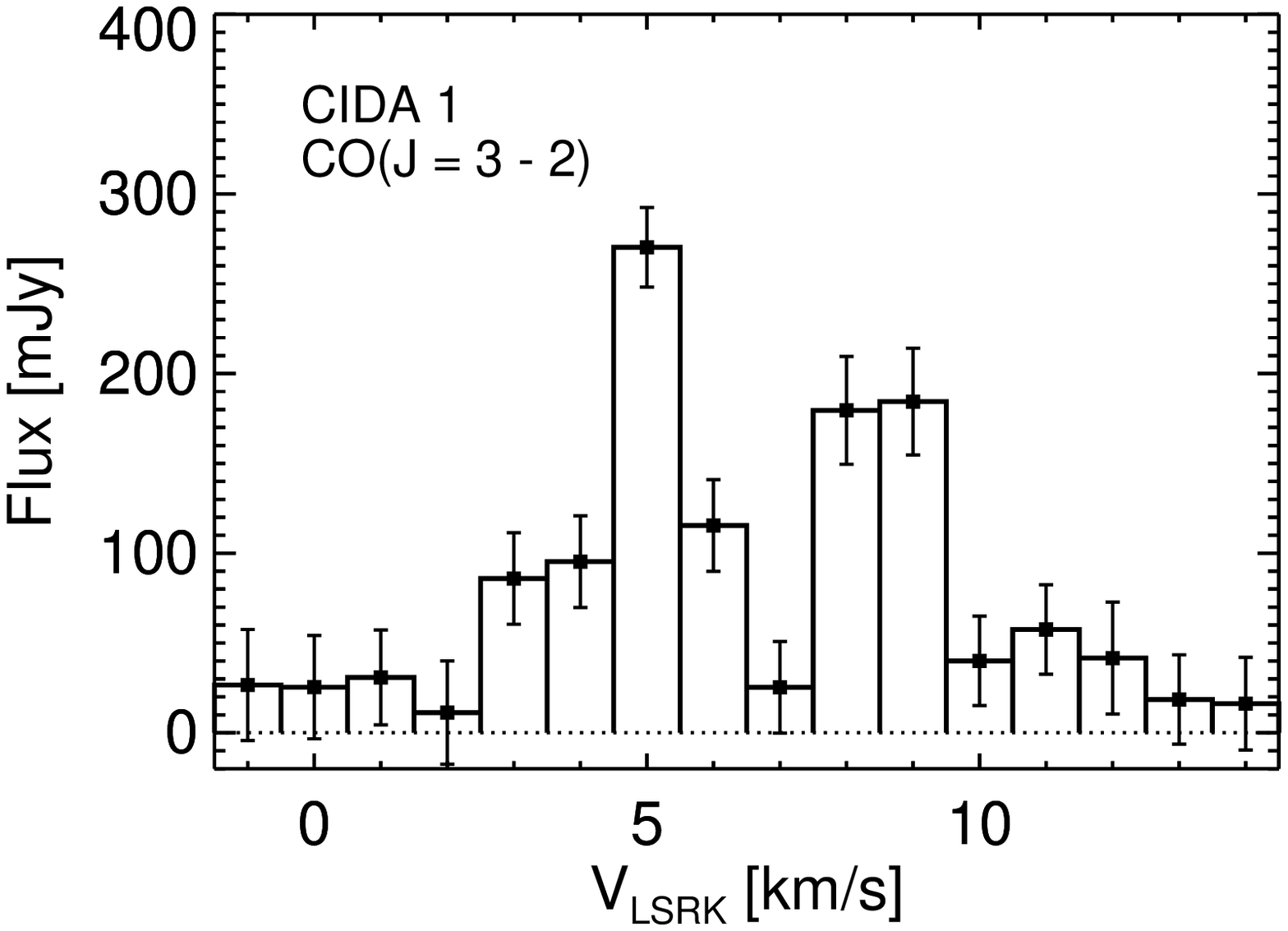}
\end{tabular}
\caption{Spatially integrated CO($J = 3 - 2$) spectra for 2M0444 (left) and CIDA 1 (right). Errorbars on flux values represent 3 times the noise level measured in each channel.}
\label{fig:co_spectra}
\end{figure*}

\begin{figure*}
\centering
\begin{tabular}{cc}
\includegraphics[scale=0.55,trim=60 0 0 0]{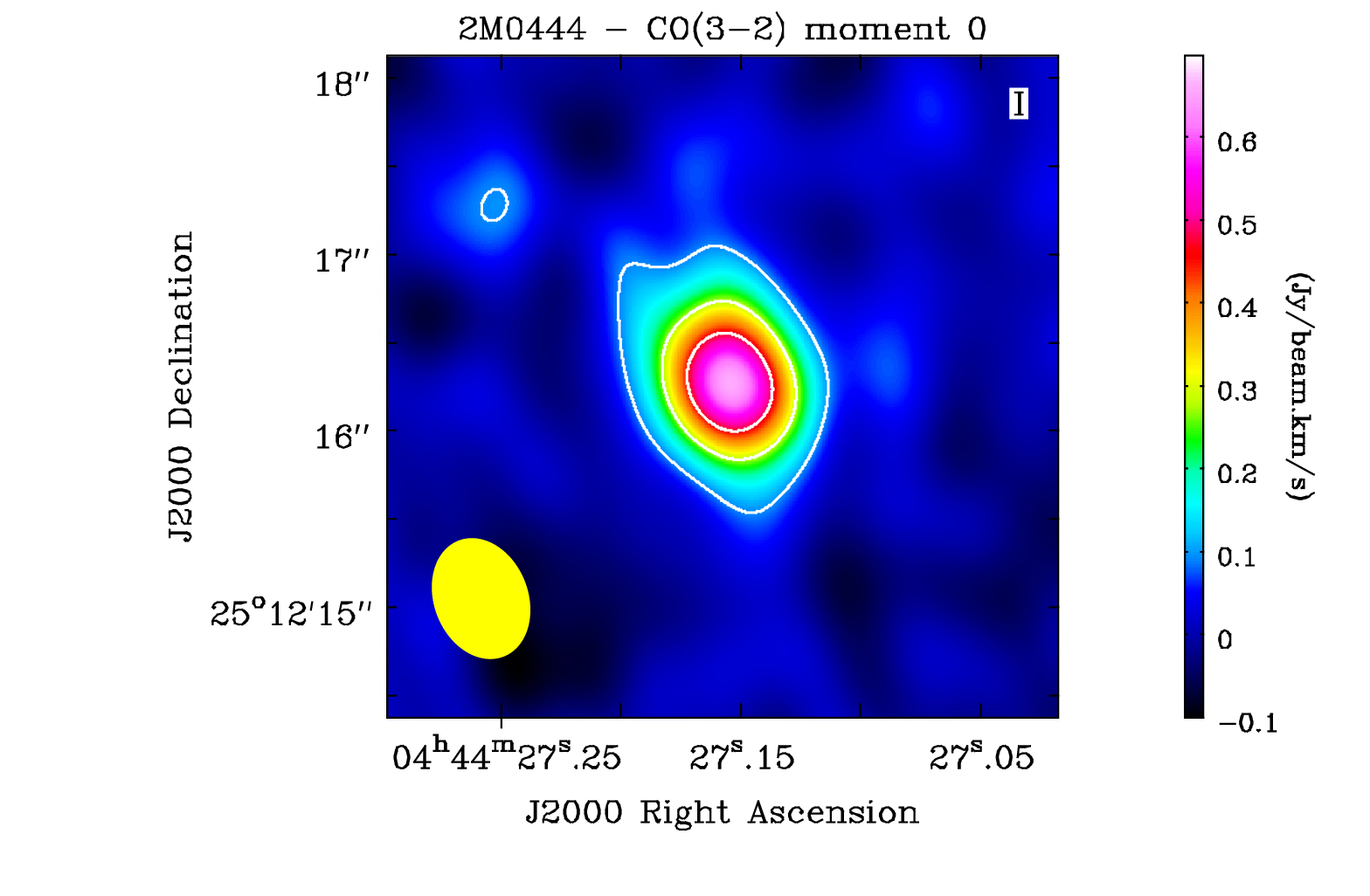} & 
\includegraphics[scale=0.5,trim=0 0 0 0]{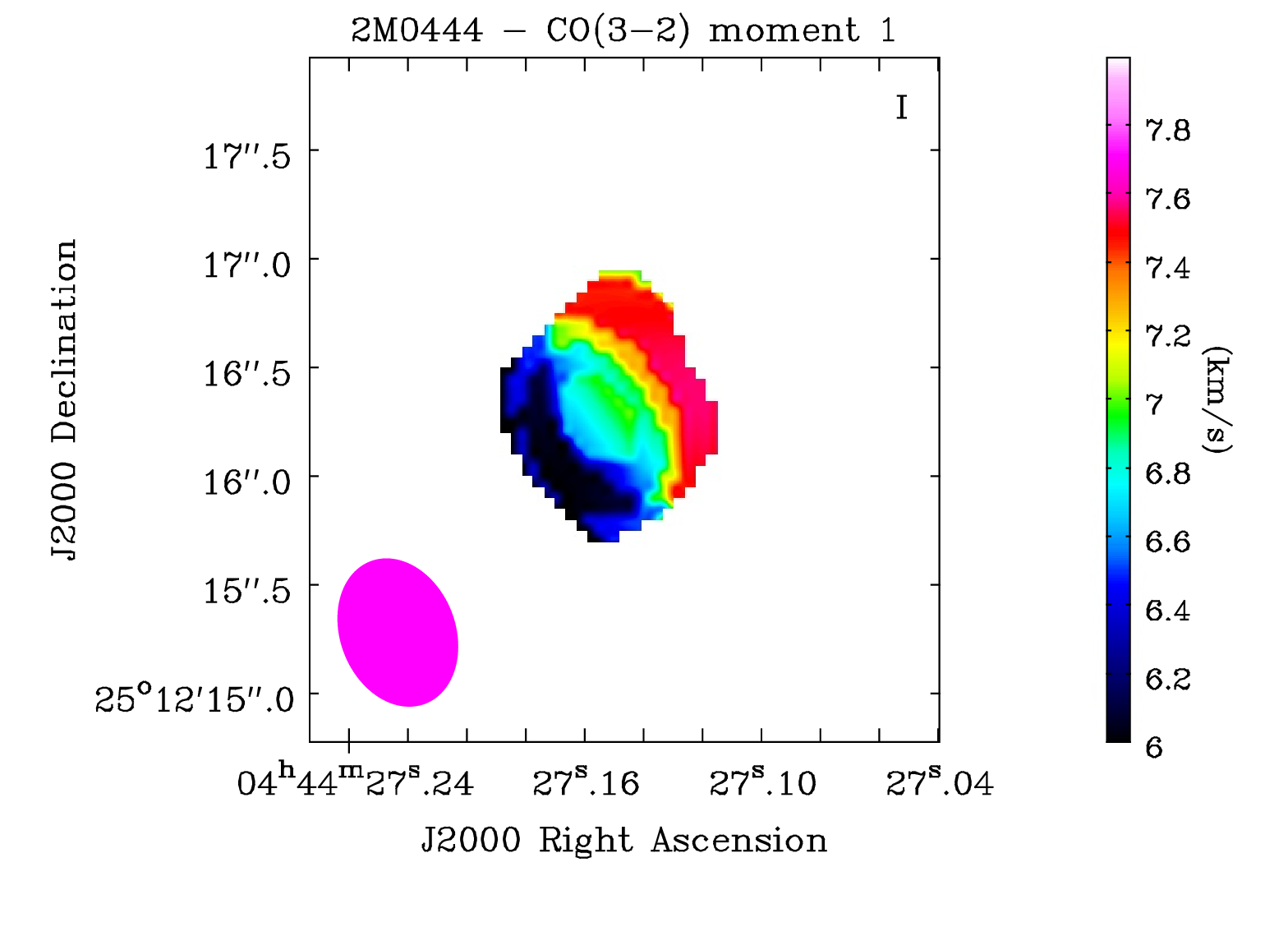}
\end{tabular}
\caption{ALMA moment 0 and 1 maps of CO($J = 3 - 2$) from 2M0444 in Band 7. Left) Moment 0 map. Contours are drawn at 3, 9, 15$\sigma$, with $1\sigma = 0.030$~Jy/beam$\cdot$km/s. Right) Moment 1 map. Only pixels with values greater than about 5$\sigma$ were used for the map. In both panels the elliptical ellipse in the lower left corner shows the synthesized beam, with sizes of $0.70'' \times 0.52''$ and P.A. $= 21.2$ degrees.} 
\label{fig:2m0444_co_mom}
\end{figure*}

\begin{figure*}
\centering
\begin{tabular}{cc}
\includegraphics[scale=0.66,trim=100 0 0 0]{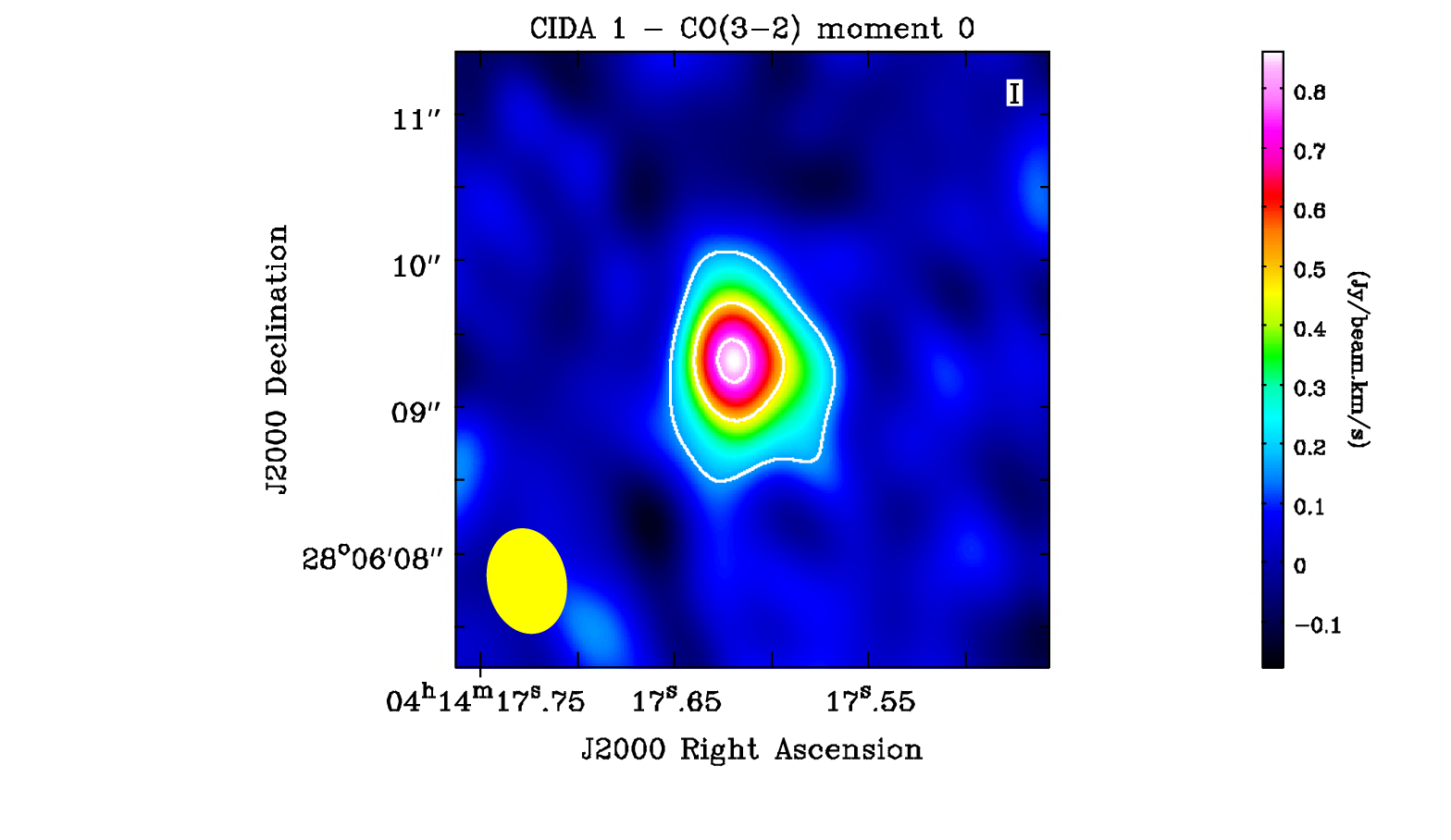} & 
\includegraphics[scale=0.44,trim=30 0 0 0]{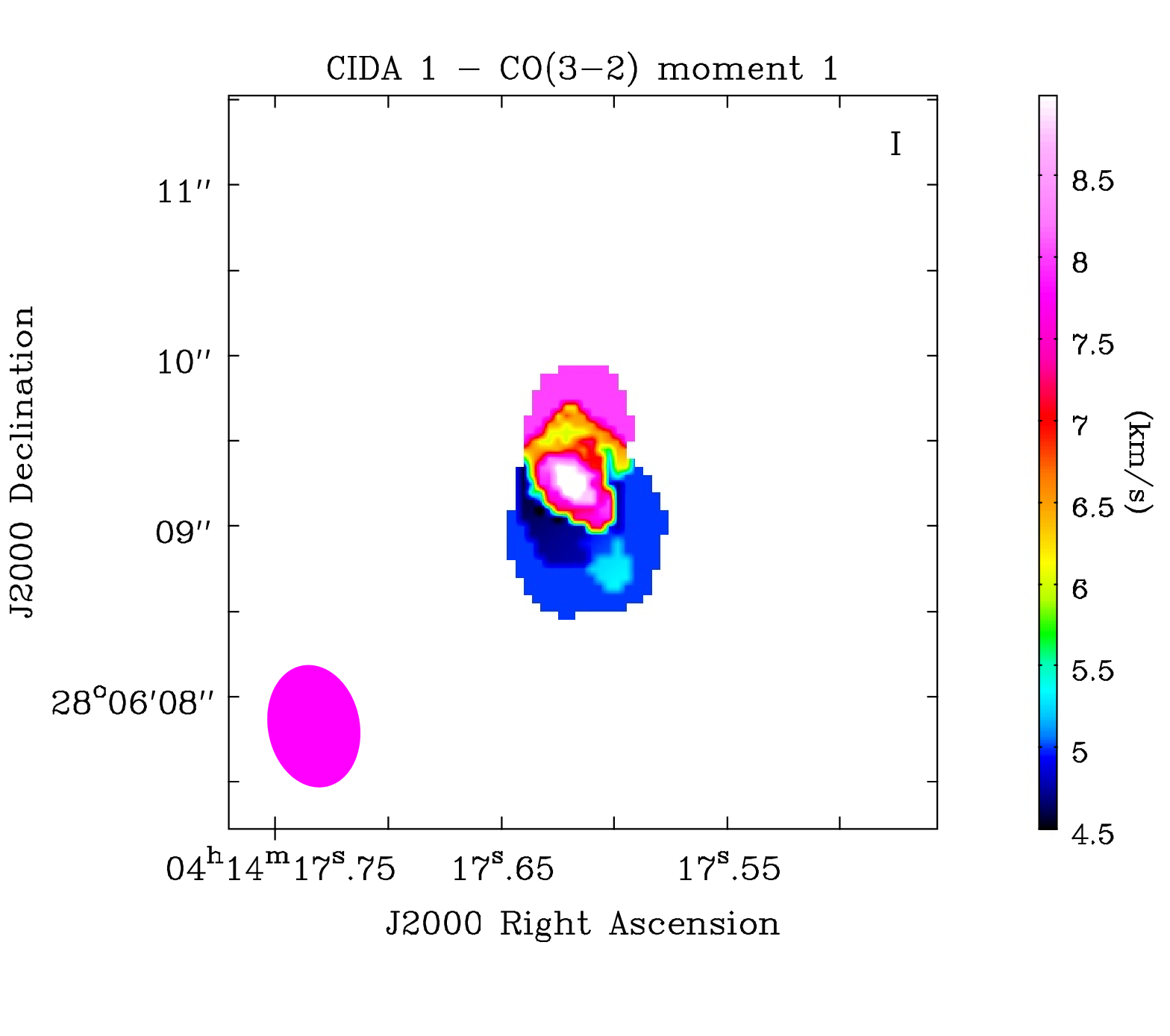}
\end{tabular}
\caption{ALMA moment 0 and 1 maps of CO($J = 3 - 2$) from CIDA 1 in Band 7. Left) Moment 0 map. Contours are drawn at 3, 9, 15$\sigma$, with $1\sigma = 0.053$~Jy/beam$\cdot$km/s. Right) Moment 1 map. Only pixels with values greater than about 5$\sigma$ were used for the map. In both panels the elliptical ellipse in the lower left corner shows the synthesized beam, with sizes of $0.72'' \times 0.53''$ and P.A. $= 11.2$ degrees.} 
\label{fig:cida1_co_mom}
\end{figure*}

\noindent Emission from the $J=3-2$ rotational transition of molecular CO was detected in Band 7 for two disks, 2M0444 and CIDA 1, and not detected for CFHT Tau 4. Figure~\ref{fig:co_spectra} shows the integrated spectra of CO$(J=3-2)$ for the two detected sources.
Figures~\ref{fig:2m0444_co_mom} and \ref{fig:cida1_co_mom} present the moment 0 and 1 maps of CO$(J=3-2)$ from 2M0444 and CIDA 1, respectively. 

Moment 0 maps quantify the source intensity integrated over the spectral extent of the line.
From these maps we integrated the surface brightness in pixels with emission from each source and derived flux values of $0.98 \pm 0.03$ and $1.45 \pm 0.05$ Jy$\cdot$km/s for 2M0444 and CIDA 1, respectively, where uncertainties reflect only the rms noise on the maps. In the case of CIDA 1, however, the CO spectrum appears to be contaminated by absorption from the foreground molecular cloud (see the central frequency  channels on the right panel in Fig.~\ref{fig:co_spectra}). If that is the case, then our flux estimate would be a lower limit.

Moment 1 maps visualize the intensity-weighted line-of-sight component of the velocity field of the source.   
The moment 1 maps of 2M0444 and CIDA 1 show a velocity gradient across both sources. In the case of the 2M0444 disk the velocity gradient is seen at a position angle of $\sim 135$ degrees (position angle is defined east of north). In CIDA 1, the moment 1 map is less ordered than in the case of 2M0444 but a velocity gradient along the north-south direction is still visible. Velocity gradients in circumstellar disks are usually dominated by rotation of the disk regulated by the gravitational potential of the central star \citep[e.g.][]{Dutrey:1998}. 
A gradient along the major axis is expected from rotation in a disk, the exact form of the radial velocity law cannot be estimated with the current data.


At the same time, the moment 1 maps are useful for the interpretation of the interferometric visibilities as they provide constraints to the geometry of the disk, i.e. disk inclination and position angle, as described in the next Section.

\begin{table*}
\centering \caption{Results of the ALMA continuum observations. } \vskip 0.1cm
\begin{tabular}{lccccc}
\hline
\hline

\\ 
Source      &   $F_{\rm{\nu=338~GHz}}$   & rms$_{\rm{\nu=338~GHz}}$ &   $F_{\rm{\nu=93~GHz}}$   & rms$_{\rm{\nu=93~GHz}}$ & $\alpha_{\rm{338-93 GHz}}$     \vspace{1mm} \\
     &        (mJy)                        		&  (mJy/beam)                           &        (mJy)                     &   (mJy/beam)   &  \\
  
\\
\hline
\\
2M0444                   & 9.0          &       0.2        &     0.87  & 0.03   & 1.81 $\pm$ 0.11 \vspace{1mm} \\
CIDA 1                     & 27.0        &       0.3       &      2.1    & 0.03   & 1.98 $\pm$ 0.11 \vspace{1mm} \\
CFHT Tau 4            & 4.3          &       0.2       &      0.31  &  0.02  & 2.04 $\pm$ 0.13  \vspace{1mm} \\

\\
\hline
\end{tabular}

\label{tab:fluxes}

\end{table*}

\section{Analysis of the interferometric visibilities}
\label{sec:visibilities}

\noindent In order to better characterize the physical extent of the disks and investigate the radial structure of the dust, a quantitative interpretation of the visibility data using physical models of disks is needed. We adopt a similar procedure as outlined in \citet{Isella:2009} and \citet{Ricci:2013}. While these papers can be used as references to the method used for our analysis, in the following we provide a brief summary of the salient points.

We consider two-layer (surface$+$midplane) models of flared disks in vertical hydrostatic equilibrium and heated by the radiation of the central star \citep{Chiang:1997,Dullemond:2001}. 
The stellar properties (effective temperature, luminosity and mass) considered for our analysis are listed in Table~\ref{tab:stellar_properties}. Dust opacities are derived by using the same method as described in \citet{Ricci:2010a}. The dust grain is modeled as a porous composite sphere made of astronomical silicates, carbonaceous materials, and water ice \citep[optical constants from][respectively]{Weingartner:2001,Zubko:1996,Warren:1984} and used a simplified version of the relative abundances suggested by \citet[][]{Pollack:1994}, as done in \citet[][]{Ricci:2010a,Ricci:2010b}. The grain size distribution in the disk midplane, which dominates the thermal emission at sub-mm/mm-wavelengths, is assumed to be proportional to a power-law $n(a) \propto a^3$ between a minimum grain size of 0.1 $\mu$m and a maximum grain size which is consistent with the sub-mm/mm spectral index as described in Section~\ref{sec:discussion}. The choice of the value of 3 for the power-law index was dictated by the low values of the dust opacity spectral index $\beta$ inferred by our analysis and presented in Section~5.1. An ISM-like value of 3.5, corresponding to a grain size distribution skewed towards smaller grains, would not be able to reproduce these values of $\beta$~\citep[see][for a discussion]{Ricci:2010a}. Smaller grains with sizes of $\sim 0.1~\mu$m are considered in the disk surface layers. Although this method provides a relatively simple recipe to evaluate dust opacities from a physical model of the dust grains, the absolute values of dust opacities in real disks still remain largely unconstrained. At the same time, only a very few results presented here are affected by this limited knowledge and are discussed where appropriate.   

The disk structure is assumed to be axisymmetric, with a surface density radial profile treated as a power-law $\Sigma(r) = \Sigma_{10 \rm{AU}} (R/\rm{10\,AU})^{-p}$ from an inner radius $R_{\rm{in}}$ to an outer radius $R_{\rm{out}}$. For the disk inner radius $R_{\rm{in}}$ we considered a value of 0.05~AU, which is approximately the dust sublimation radius given the (sub-)stellar properties of the sources considered in this work. The exact value of $R_{\rm{in}}$ does not affect the analysis of the millimeter visibilities, which are dominated by emission from regions at much larger radii. 

\begin{figure}
\centering
\includegraphics[scale=0.45]{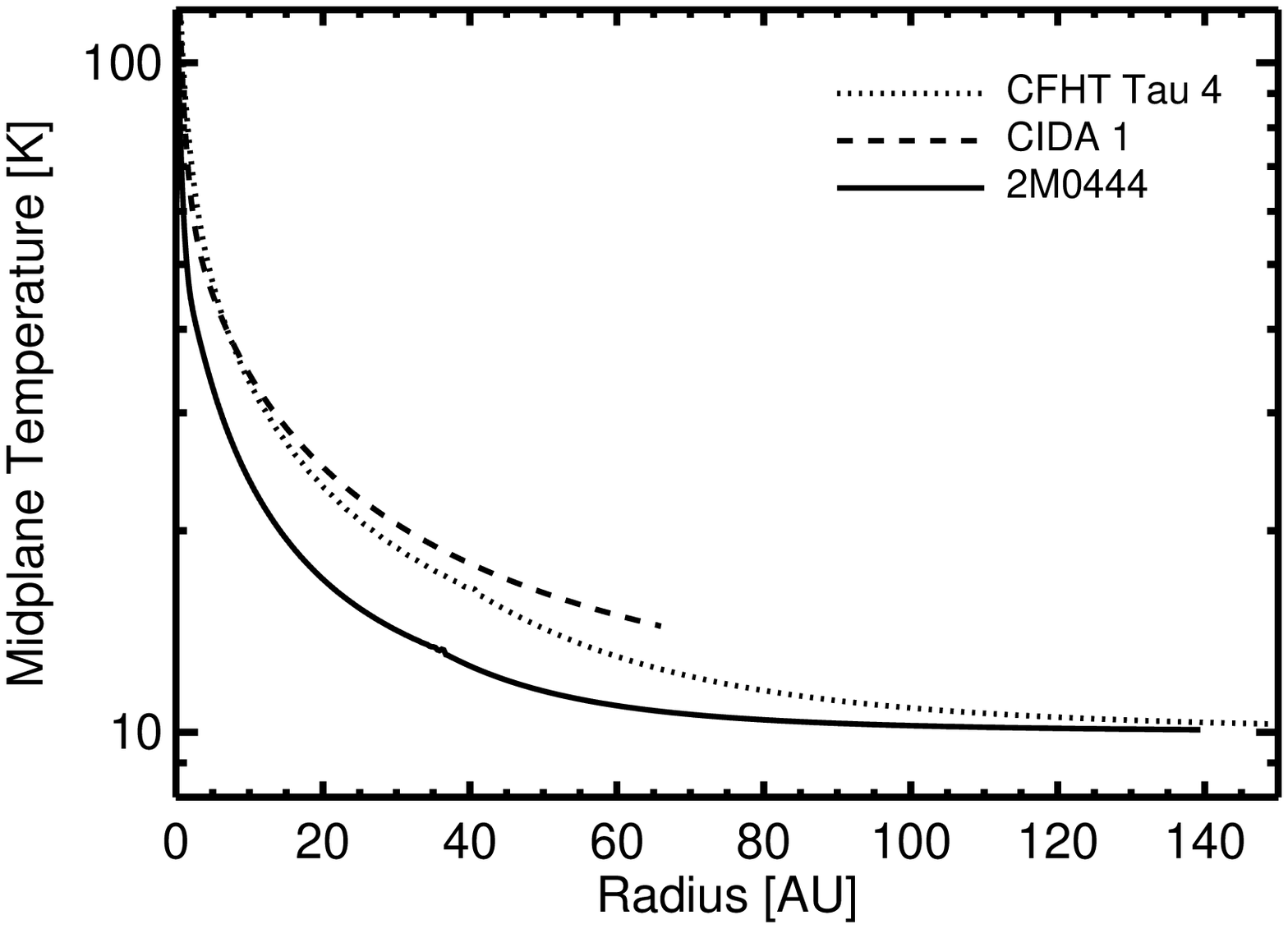}
\caption{Radial profile of the dust temperature in the disk midplane for the best fit models presented in Section~4.}
\label{fig:temp}
\end{figure}

We note here that for a significant fraction of the models considered in this work, the surface of the disk is optically thin to the stellar radiation at a radius $R_{\rm{thin}}$ smaller than $R_{\rm{out}}$. The exact value of $R_{\rm{thin}}$ depends on both the stellar properties, which determine the stellar spectrum impinging the disk, and also the disk physical structure. For example, in the case of 2M0444 and CFHT Tau 4, the best fit models have a $R_{\rm{thin}}$ value of $\approx$ 40~AU, while $R_{\rm{out}} > 100$ AU (see below); instead, in the case of CIDA 1 the best fit model is optically thick to the stellar radiation at all disk radii.
In the standard formulation of the two-layer disk models, the disk would be truncated at $R_{\rm{thin}}$ as beyond this radius the disk surface, one of the two ``layers'' of these models, would not be defined.  

Since the ALMA visibilities in Band 7 indicate that the 2M0444 and CFHT Tau 4 disks have thermal emission beyond $R_{\rm{thin}}$, we extend the temperature profile of the disk models using a power law with an index of -3/4 between $R_{\rm{thin}}$ and $R_{\rm{out}}$, and we also include a contribution of heating from the diffuse interstellar radiation field to avoid unreasonably low temperatures in the disk outer regions. Therefore, while for disk radii smaller than $R_{\rm{thin}}$ the temperature of the disk is calculated by the two-layer models, at radii larger than $R_{\rm{thin}}$ the disk temperature is given by 
$T^4(R) =(T(R_{\rm{thin}})\cdot(R/R_{\rm{thin}})^{-3/4})^4+T^4_{\rm{ext}}$, where $T_{\rm{ext}}$ describes the effect of any external heating source. We chose a fiducial value for $T_{\rm{ext}} = 10$ K, as observed in the outermost regions of dense cores directly heated by the diffuse interstellar radiation in molecular clouds \citep[e.g.][in the case of the TMC-1C core in Taurus]{Schnee:2005}. 

Figure~\ref{fig:temp} shows the radial profiles of the dust temperature for the best fit models presented below. In the case of 2M0444 and CFHT Tau 4, the radial profile gets shallower in the outermost regions and approaches $T_{\rm{ext}}$ asymptotically. In our analysis
we also considered values of 5 and 15~K for $T_{\rm{ext}}$, and found best fit models with parameters which lie within the confidence intervals obtained in the 10 K case. This shows that, although arbitrary, the choice of this value does not have a significant impact on the main conclusions of this work. 
As for the radial index of the stellar heating beyond $R_{\rm{thin}}$, our choice of -3/4, lower than the value of $\approx -1/2$ for a flared disk which efficiently absorbs stellar photons, was motivated by the less efficient heating expected beyond $R_{\rm{thin}}$ because of the lower optical depths ($< 1$) to stellar radiation. A more realistic estimate requires sophisticated radiative transfer calculations. However, what is important to notice for our analysis is that, for both the best fit models for 2M0444 and CFHT Tau 4, $T(R_{\rm{thin}})$ turns out to be very close to $T_{\rm{ex}}$. Therefore, the temperature profiles for these two disks become quickly dominated by the $T^4_{\rm{ex}}$ term, as shown in Fig.~\ref{fig:temp} in the $T_{\rm{ex}} = 10$ K case, and the main results of our analysis are nearly insensitive to the specific choice to describe the stellar heating in that region.


The free model parameters for the disk structure are $\Sigma_{\rm{10AU}}$, $p$, $R_{\rm{out}}$, or, equivalently, $M_{\rm{disk}}$, $p$, $R_{\rm{out}}$, as there is a one-to-one relation between $\Sigma_{\rm{10AU}}$ and the disk mass $M_{\rm{disk}}$ once $p$, $R_{\rm{out}}$ (and $R_{\rm{in}}$) are specified. For 2M0444 and CIDA 1 we considered intervals for the disk inclination $i$ (angle between the disk rotation axis and line-of-sight), and position angle (P.A., angle between North and disk semi-major axis toward East) from the CO moment 1 maps presented in Section~\ref{sec:co}. In particular, the P.A. of the disk was estimated from the angle of the axis joining the blue-shifted pixels to the red-shifted pixels in the moment 1 maps. A reasonable range of possible values is considered to account for the uncertainties given by the limited angular and spectral resolution of our observations. Wide intervals for the disks inclination were considered also because of the relatively elongated synthesized beams which do not allow very accurate estimates of the disks aspect ratio (see Figures \ref{fig:2m0444_co_mom},\ref{fig:cida1_co_mom}).
These intervals are listed in Table~\ref{tab:disk_properties}. For CFHT Tau 4 for which CO was not detected, the disk inclination and P.A. were left free to vary and constrained by the fitting procedure.

These models calculate the predicted map of the disk surface brightness at any given frequency, and model visibilities are derived calculating its Fourier transform.
The model visibilities are then interpolated on the grid of $(u,v)$-points sampled by the ALMA observations in Band 7, and compared with the measured visibilities using a $\chi^2$-minimization algorithm.   
We used a Markov Chain Monte Carlo (MCMC) to run large collections of disk models by varying the input parameters that our analysis aims to constrain. The computed $\chi^2$-values are then used to derive the confidence intervals for the constrained input parameters of the disk models.   

\begin{figure*}
\centering
\begin{tabular}{cc}
\includegraphics[scale=0.5,trim=0 0 0 0]{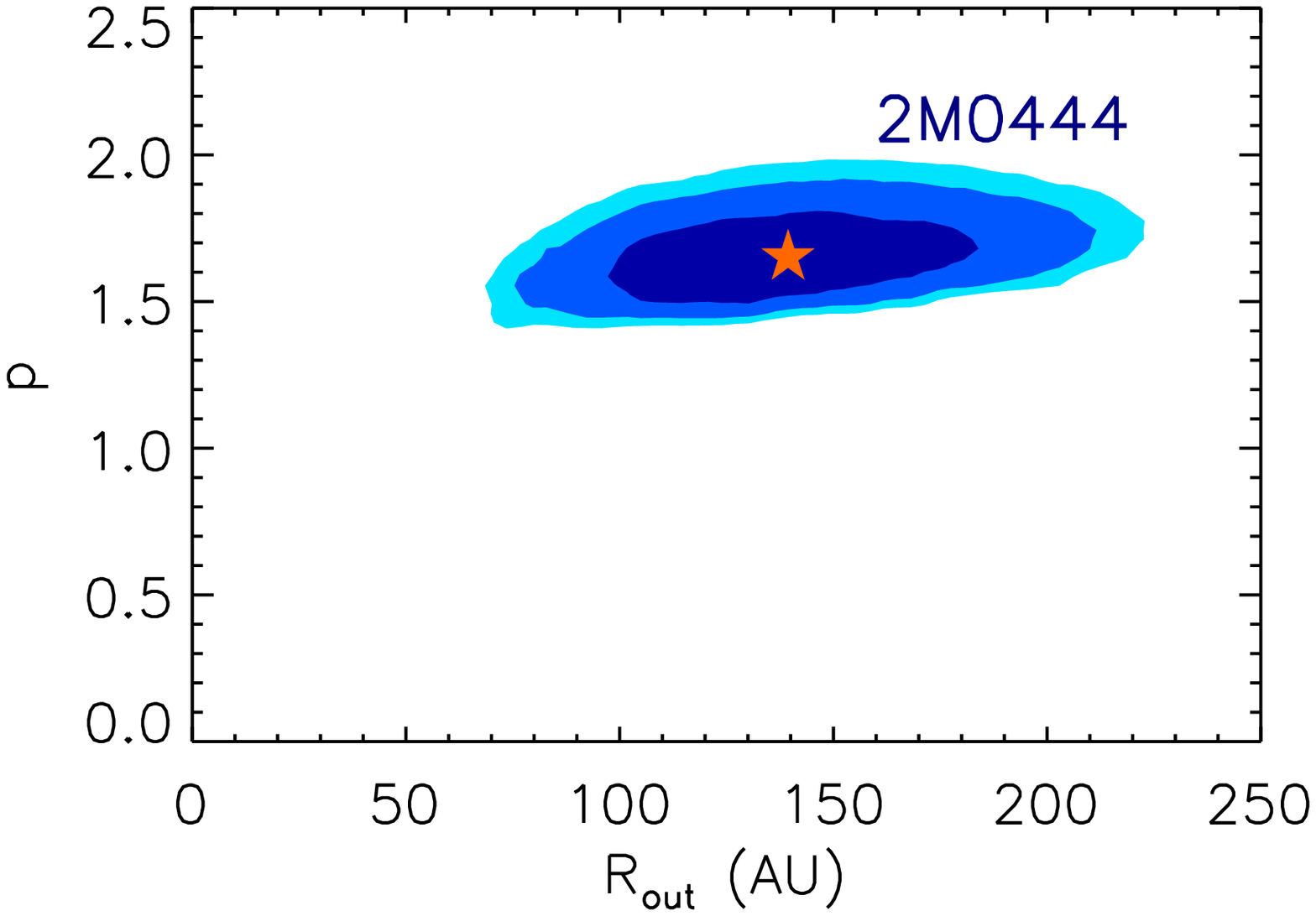} & 
\includegraphics[scale=0.5,trim=0 0 0 0]{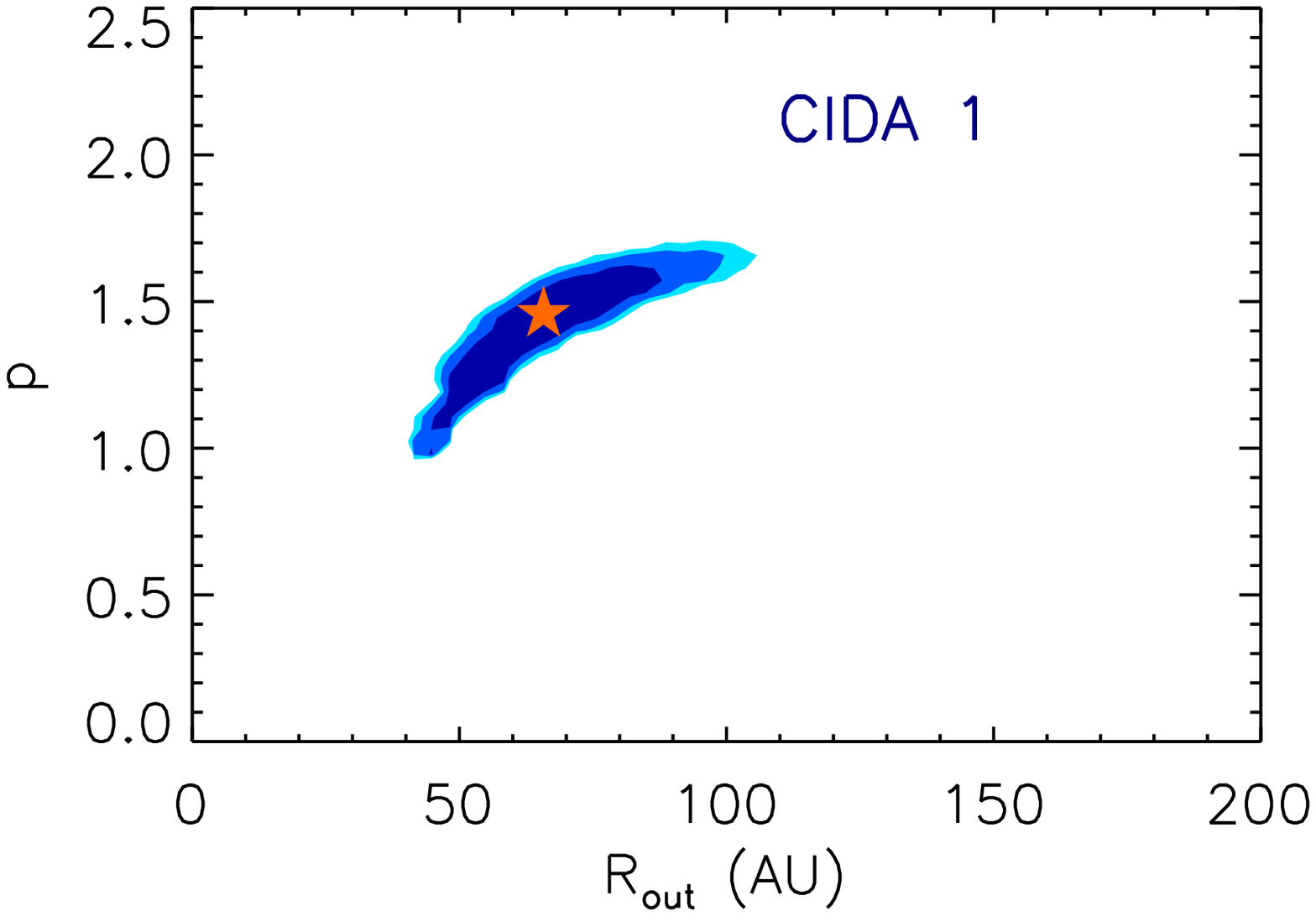} \\
 \multicolumn{2}{c}{\includegraphics[scale=0.5,trim=0 0 0 0]{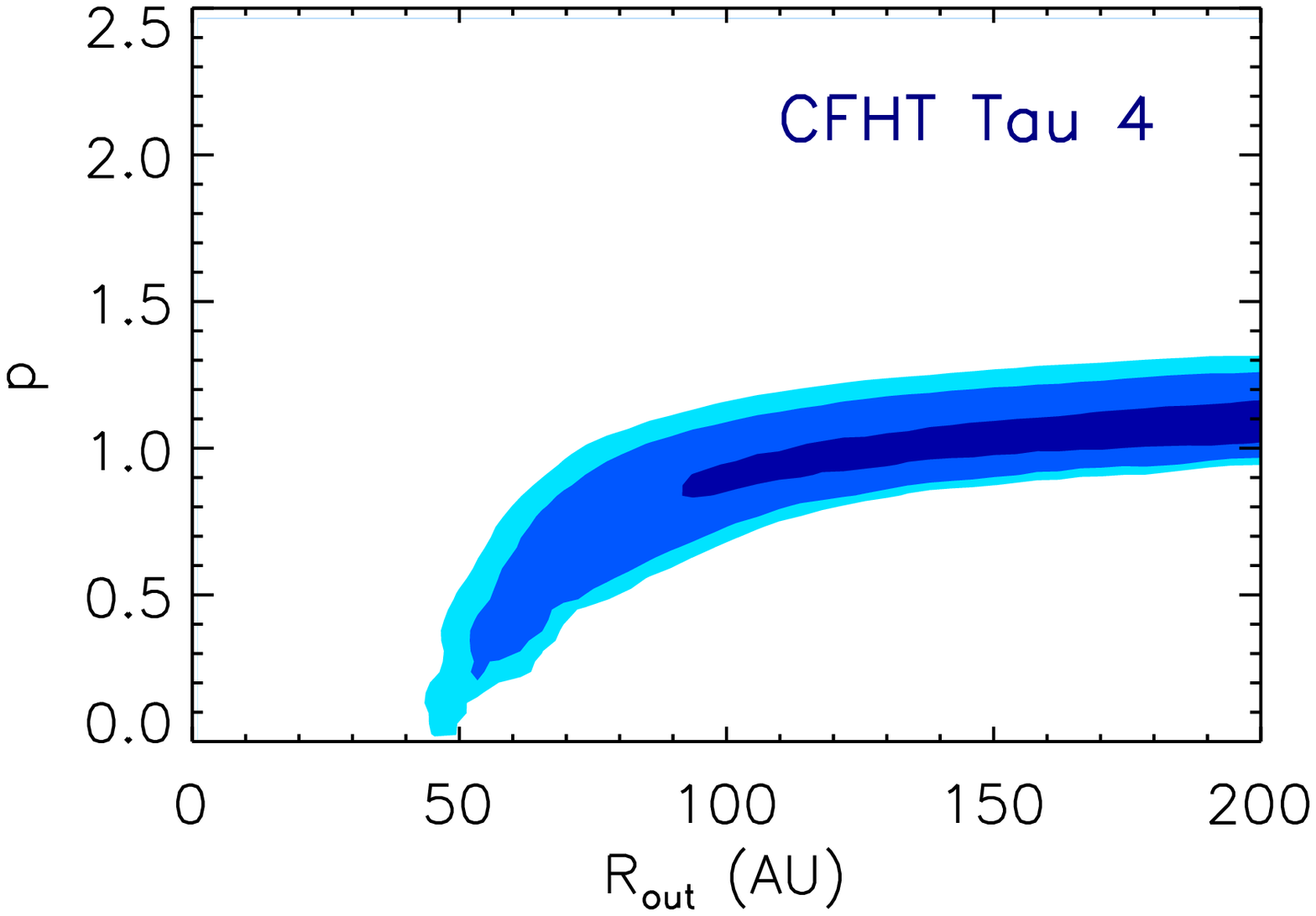}}
\end{tabular}
\caption{2D contour plots of the $\chi^2$-function projected over disk outer radius $R_{\rm{out}}$ and slope $p$ of the radial profile of the dust surface density and derived from the comparison between two-layer disk models and ALMA Band 7 data of 2M0444 (top left panel), CIDA 1 (top right) and CFHT Tau 4 (bottow). Contours with different tones of blue are drawn at the $68\%, 90\%, 95\%$ 2D confidence intervals. The orange stars represent the best-fit disk models. In the case of CFHT Tau 4 the star is not shown because only a lower limit for $R_{\rm{out}}$ could be obtained and the outer disk radius of the nominal best-fit model lies outside the range shown in the plot.} 
\label{fig:chi_2m0444}
\end{figure*}

Figure~\ref{fig:amp_vis} shows the real parts of the visibility function of our best-fit disk models binned over deprojected baseline length and over plotted to the ALMA data in Band 7 for each disk. 
Figure~\ref{fig:chi_2m0444} shows the 2D contour plots of the $\chi^2$ function over the ($R_{\rm{out}}$, $p$) parameter space.
The constraints on the disk parameters derived by our analysis are listed in 
Table~\ref{tab:disk_properties}. The values, or lower limits, obtained for the outer disk radii of all the three disks are larger than the spatial resolution of our ALMA Band 7 observations, i.e. about 28~AU in radius. This confirms that our observations have spatially resolved the dust emission of all the disks presented here. 
In the case of CFHT Tau 4, only a lower limit for $R_{\rm{out}}$ could be derived. This is because the disk surface brightness rapidly decreases at large distances from the central object. 

\begin{table*}
\centering \caption{Summary of the stellar properties used for the analysis. } \vskip 0.1cm
\begin{tabular}{lcccc}
\hline
\hline

\\ 
Source      &  SpT   & $T_{\rm{eff}}$  &  $L_{\star}$  & $M_{\star}$      \vspace{1mm} \\
     &                     &  (K)                           &     ($L_{\sun}$)                       &   ($M_{\sun}$)    \\
  
\\
\hline
\\
2M0444                   & M7.25    &       2838        &     0.028  & 0.05    \vspace{1mm} \\
CIDA 1                     & M5.5      &       3058        &     0.126  & 0.09    \vspace{1mm} \\
CFHT Tau 4            & M7         &       2880        &      0.175  & 0.095   \vspace{1mm} \\

\\
\hline
\end{tabular}

\begin{flushleft}
\textbf{Notes.} Spectral types are from \citet{Luhman:2010}. The derivation of the other stellar parameters is described in \citet{Ricci:2013} for 2M0444 and \citet{Andrews:2013} for CIDA 1 and CFHT Tau 4.
\end{flushleft}

\label{tab:stellar_properties}

\end{table*}

%
%

\begin{table*}
\centering \caption{Disk parameters constrained by our analysis. } \vskip 0.1cm
\begin{tabular}{lccccc}
\hline
\hline

\\ 
Source    & $R_{\rm{out}}$  &  $p$  & P.A.      & $i$ & $M_{\rm{disk}} \times \frac{\kappa_{\rm{0.89mm}}}{0.02~\rm{cm^2/g_{gas}}}$\vspace{1mm} \\
                         &  (AU)                           &                        & (degrees) & (degrees)  & ($M_{\rm{Jup}}$)    \\
  
\\
\hline
\\
2M0444                           &         139 $^{+20}_{-27}$     &    1.65 $^{+0.07}_{-0.10}$   &   100 - 130     &  30 - 60 & 1.3 $^{+0.2}_{-0.2}$ \vspace{2mm} \\
CIDA 1                            &   66 $^{+14}_{-12}$     &   1.46  $^{+0.13}_{-0.21}$        &  -10 - 40 & 30 - 70 & 2.1 $^{+0.2}_{-0.4}$  \vspace{2mm} \\
CFHT Tau 4            &        $>$ 80       &   1.20 $^{+0.10}_{-0.45}$    &   20 - 30     &  75 - 80  & 0.8 $^{+0.2}_{-0.4}$ \vspace{2mm} \\

\\
\hline
\end{tabular}

\begin{flushleft}
\textbf{Notes.} Intervals for each parameter are given at the $1\sigma$ confidence level. The upper limit of 80 degrees for the interval of the inclination of the CFHT Tau 4 disk is given by the fact that a disk with $i~\simgreat$ 80~degrees would strongly occult the central object \citep[e.g.][]{Skemer:2011}. For 2M0444 and CIDA 1, intervals for position angle and disk inclination are from the CO momentum 1 map; for CFHT Tau 4 the intervals come from the fitting of the dust continuum visibilities.
Estimates for the disk masses are given relative to a value of 0.02~cm$^2/$g$_{\rm{gas}}$, derived from the dust model described in Section~\ref{sec:visibilities}. The ISM-like value of 100 for the gas-to-dust mass ratio is also assumed. 
\end{flushleft}
\label{tab:disk_properties}

\end{table*}

\section{Discussion}
\label{sec:discussion}

\subsection{Large grains in disks around VLMs/BDs}

\noindent All the three disks show very similar values of their spectral index, which are consistent to each other within 1$\sigma$.  These values, $\approx 1.8 - 2.0$, are significantly lower than the typical values of $\approx 3.5 - 4.0$ measured at these frequencies for dust in ISM diffuse regions and HII regions in the Galaxy \citep[e.g.][and references therein]{Planck-Collaboration:2013}. If the dust emission probed by the ALMA continuum observations is optically thin and in the Rayleigh-Jeans tail of the spectrum, the spectral index $\alpha$ directly traces the spectral index of the dust opacity $\beta$ ($\kappa_{\nu} \propto \nu^{\beta}$), with the two indices tied by the relation $\beta = \alpha - 2$. As the $\beta$ index is related to the grain size of the emitting dust \citep{Natta:2007}, with $\beta < 1$ (or $\alpha < 3$) interpreted as evidence for mm-sized grains or larger and $\beta \approx 1.5 - 2.0$ consistent with smaller sub-$\mu$m sized dust, the low values of $\alpha$ measured for our three disks would imply the presence of grains with sizes larger than about 1 mm in the disk.

This interpretation of mm-sized grains to explain the low values of the spectral index $\alpha$ can be modified in the case either one or both the assumptions mentioned above, i.e. optically thin emission in the Rayleigh-Jeans regime, are incorrect. If the disk is small enough, the bulk of the dust emission would be produced in the inner regions of the disk. These regions can be so dense that the optical depth can approach values of the order of unity even at millimeter wavelengths, and this has the effect of making the spectrum look shallower at these frequencies.  In the extreme case of an optically thick disk ($\tau_{\nu} >> 1$), the dust emission becomes insensitive to the dust opacity coefficient, and therefore no information on the $\beta$ index, and dust grain sizes, can be extracted \citep[][]{Testi:2001}. Also, if the temperature of the emitting dust is cold enough the Planck function cannot be approximated by a power-law in frequency with spectral index of 2 as in the Rayleigh-Jeans regime and its frequency dependence is shallower. As a consequence, the value of the spectral index of the SED is lower than in the Rayleigh-Jeans approximation. 

\citet{Ricci:2012a,Ricci:2013} investigated these effects in the disks around $\rho-$Oph 102, an object with a M5.5-M6 spectral type with an age of $\sim 1$ Myr, similar to the objects presented in this paper, and 2M0444, which is one of the three disks discussed here. They found that the measured sub-mm/mm fluxes can be interpreted by two different classes of disk models. Compact optically thick disks, with outer radii $\approx 5 - 10$ AU and unconstrained dust properties, as well as larger disks containing dust grains with sizes of at least $\sim 1$ mm can explain the sub-mm/mm photometry at the same time.
The only direct way to discriminate between these two alternative hypotheses and investigate the grain size of the dust is to spatially resolve the dust emission \citep[][]{Testi:2003,Ricci:2013}. 

The results of the analysis described in Section~\ref{sec:visibilities} indicate that the low values of the sub-mm/mm spectral index measured for the three disks cannot be explained by small, $R_{\rm{out}} \simless 10$ AU, optically thick disks \citep{Ricci:2012a,Ricci:2013}. 
The only way to reproduce the ALMA fluxes is through spectral indices of the dust emissivity $\beta < 1$. Our models constrain $\beta = 0.2 \pm 0.2$, $0.3 \pm 0.2$, $0.4 \pm 0.2$ for 2M0444, CIDA 1, CFHT Tau 4, respectively.   
The fact that $\alpha - \beta < 2$ is because the outer regions of disks around these VLMs and BDs are relatively cold, the most extreme case being 2M0444 for which our models predict a dust temperature $T_{\rm{dust}} \approx 12$ K at a distance of about 40~AU from the brown dwarf (see Fig.~\ref{fig:temp}). At these low temperatures the dust emission is not in the Rayleigh-Jeans regime and therefore the spectrum is shallower than the Rayleigh-Jeans case at a given $\beta$.  

These values of $\beta$ indicate that in all the three disks dust grains as large as at least $\sim 1$ mm are present in the disk. In particular, the dust model described in Section~\ref{sec:visibilities} predicts maximum grain sizes $a_{\rm{max}} \sim 1 - 10$ cm to reproduce the derived $\beta$-values. Similar values of $a_{\rm{max}}$ are obtained if using the dust models made of astronomical silicates, troilite, organic materials and water ice presented in \citet{D'Alessio:2001}, while larger values would come by dust made of olivine, organics and water ice considered in \citet{Natta:2007}. However, we want to emphasize that the quantity inferred by our analysis is the spectral index $\beta$ of the dust opacity law, whereas the relation between $\beta$ and grain sizes strongly depends on the adopted dust model. This is because the slope of the opacity law at these wavelengths depends on grain properties, such as chemical composition and morphology, which are highly unknown for grains in the midplane of real disks~\citep[see e. g. the discussions in ][]{Natta:2004, Draine:2006, Banzatti:2011}. The robust statement to be considered here is that for our inferred values of $\beta < 1$ grains with sizes of the order of $\sim 1$ mm or larger need to be invoked \citep[][]{Draine:2006}.

Although our observations do not allow to pinpoint the exact location of these large grains in the disks, it is clear that the sub-mm/mm spectral index mostly probes dust in the cold outer disk regions, where the bulk of the (spatially unresolved) emission at the frequencies probed by our ALMA observations originates from.
Recent works which have constrained the spatial variation of $\beta$ across the disk via multi-wavelength observations at high angular resolution of T Tauri stars have found a general radial trend, with lower $\beta$-values (larger grains) closer to the star \citep{Guilloteau:2011,Perez:2012,Trotta:2013}. In these cases the spatially unresolved $\beta$-values are found to be good representations of the $\beta(R)$ function at stellocentric radii $R \approx 0.5 \times R_{\rm{out}}$.   

The main result of our observations is that in all the three disks around VLMs and BDs we found evidence for grains with sizes of $\sim 1$ mm or larger in the disk outer regions. This is consistent with what \citet{Ricci:2012a} found in the disk around $\rho$-Oph 102, a $\sim 1$ Myr, M5.5-M6 spectral type object in the Ophiuchus star forming region. 

\begin{figure*}
\centering
\includegraphics[scale=1.0,trim=0 0 0 0]{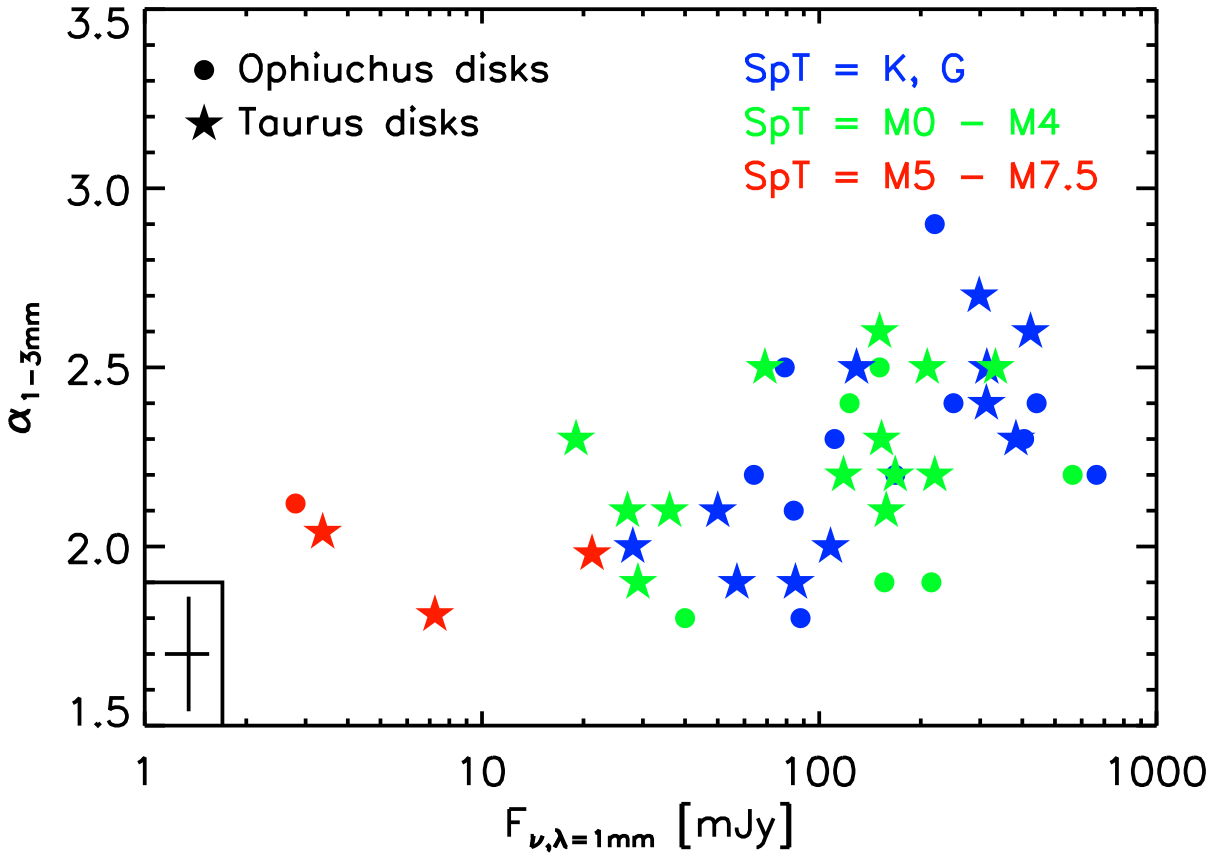}  
\caption{Flux density at 1mm vs spectral index between 1 and 3mm for a sample of disks around young PMS stars and brown dwarfs. Different colors and symbols indicate objects with different ranges of spectral types and in different star forming regions. Green and blue data points are from \citet{Ricci:2010a,Ricci:2010b,Ricci:2012b}. The three red stars are the disks presented in this work. The red circle represents $\rho-$Oph 102 after correcting the flux densities given by \citet{Ricci:2012a} for the new 2012 versions of the Butler-JPL-Horizons models for flux calibration. The updated flux densities for that object are $F_{\rm{0.89mm}} = 3.6 \pm 0.2$~mJy and $F_{\rm{3.2mm}} = 0.24 \pm 0.03$~mJy and the spectral index $\alpha_{\rm{0.89-3.2mm}} = 2.12 \pm 0.15$. The 1mm flux densities shown in the plot have been obtained by interpolating the flux densities between nearby wavelengths. The errorbars in the lower left corner show the typical uncertainties.} 
\label{fig:flux_alpha}
\end{figure*}

Figure~\ref{fig:flux_alpha} shows the mm spectral index vs flux density at 1~mm for a sample of single disks in Taurus and $\rho$-Oph 102, where we included the VLMs/BDs disks discussed here and in \citet{Ricci:2012a}. No significant trend of the spectral index $\alpha_{\rm{1-3mm}}$ is seen with spectral type. 
This suggests that grain growth to mm/cm-sized grains in the outer regions of low mass disks around VLMs/BDs is as efficient as in more massive disks in T Tauri stars.

\citet{Pinilla:2013} have investigated the evolution of solids in disks around VLMs/BDs. They calculated the time evolution of solids starting from ISM-like sub-$\mu$m sizes and accounting for coagulation, fragmentation and radial migration of solids in gas-rich disks around VLMs/BDs. They derived the time evolution of grains in disks with different physical properties \citep[varying $R_{\rm{out}}$, $p$, $\alpha_{\rm{turb}}$-parameter quantifying the level of turbulence in the disk, from][]{Shakura:1973}, different values of the fragmentation velocities which defines the collisional velocity threshold under which grains stick upon collision, and different inhomogeneities in the gas structure to mimic long-lived pressure bumps which can slow down the radial migration of solids. For each of their simulated disks, they calculated the sub-mm/mm fluxes. Among all the models that they considered, the class of models that can reproduce the sub-mm/mm fluxes of the VLMs/BDs disks presented here and in \citet{Ricci:2012a,Ricci:2013} have the following characteristics: relatively strong inhomogeneities in the pressure field of the gas in the disk to efficiently slow down the radial migration of mm/cm-sized pebbles, small disk outer radii ($R_{\rm{out}} \approx 15-30$ AU), a moderate turbulent strength ($\alpha_{\rm{turb}} < 10^{-3}$), and average fragmentation velocities for ices of about 10 m/s. 
What we want to note here is that in the case of the three disks considered in this work, the outer disk radii constrained by our analysis are significantly larger than the ones invoked by the models of \citet{Pinilla:2013}. If all else is equal, the effect of increasing the disk radius is to decrease the surface density of the disk. At lower densities the process of grain growth is less efficient and lower maximum grain sizes would be expected. To what extent this can affect the main conclusions of that work has to be explored in future modeling studies of dust evolution.

\citet{Meru:2013} considered models of solids evolution in VLMs/BDs disks with a more physical treatment for the outcome of the grains collisions. These models use a probability distribution function for the velocity of the solid particles rather than single-valued velocities and also consider the fact that collisions between low- and high-mass aggregates can withstand destruction better than collisions between equal-mass aggregates, as shown by laboratory experiments of grain collision~\citep[][]{Blum:2008,Testi:2014}. Using these models they showed that the collisional timescales for grain growth are similar for BDs and T Tauri disks under the assumption that the former are scaled down versions of the latter in terms of disk mass and radius. However, as noted previously, our observations suggest that grain growth to mm/cm grains can occur in VLMs/BDs disks with disk radii which are \textit{similar}, and not significantly smaller than T Tauri disks. Another limitation of that work is that these models neglect radial drift, which can play a crucial role in the evolution of solids \citep{Pinilla:2013}. 
Theoretical studies which include the relevant physics for the evolution of solids in disks with properties as constrained by observations such as those presented in this paper are needed to improve on our understanding of the early phases toward planet formation in low mass disks.

\subsection{The disks radial structure and the formation of brown dwarfs and very low mass stars}

\noindent Other then providing information on the efficiency of the growth of solids in disks around BDs and VLMs, our ALMA observations allow to obtain the first strong constraints on the radial distribution of dust particles in these disks. 
The values of the disk outer radius $R_{\rm{out}}$ and radial slope $p$ of the dust surface density for our disks fall well within the ranges measured for disks around T Tauri stars in the Taurus star forming region~\citep[][]{Guilloteau:2011}.

This result is consistent with all the size estimates obtained so far for disks surrounding VLMs and BDs, although less precise than the ones derived in this work \citep{Scholz:2006, Luhman:2007, Ricci:2012a,Ricci:2013}. 
These findings point toward a stellar-like scenario for the formation of these BDs and VLMs. In fact, in the main competitive theoretical framework proposed to form these low mass objects, i.e. via the ejection of pre-stellar objects in multiple systems, these disks are usually significantly smaller and with lower mass than those around more massive PMS stars \citep{Reipurth:2001}. 
Numerical simulations for the formation and early evolution of stellar clusters by \citet{Bate:2003} calculate the distribution of the outer radii for disks surrounding young BDs formed in this scenario. Because of tidal truncation in disk-star and disk-disk dynamical interactions, these disks have typical outer radii $R_{\rm{out}} \approx 10$ AU, while only for $5-10\%$ of these systems $R_{\rm{out}}$ can be larger than 20 AU. 
Somewhat larger disks are obtained by more recent hydrodynamical simulations which include radiative transfer and a more realistic equation of state \citep{Bate:2012}. 
Disk radii are computed in terms of  \textit{truncation radii} by considering the distribution of 1/2 of the closest distance among all the encounters every single object experiences during the duration of the simulation.
They find that half of the objects with masses lower than 0.1~$M_{\odot}$ have disks with truncation radii smaller than 10~AU, 20\% larger than 40 AU and $\simless 10\%$ larger than 100 AU. 

All the three disks discussed in this paper have outer radii much larger than the typical values predicted in this scenario and consistent with the values found for disks around young stars. 
This suggests the stellar-like gravitational collapse of a dense pre-(sub)stellar core as the the most plausible mechanism for the formation of these objects \citep{Andre:2012}.
However, the predicted values by the Bate simulations should be regarded as lower limits since these simulations are made to represent a dense cluster and disk truncation may be less severe in a lower-density region like Taurus. Also, some viscous spreading and further accretion from gas in the molecular cloud onto the disk could occur at ages between the end of these simulations ($\approx$ 0.3 Myr) and the typical age of disks in Taurus ($\sim 1$ Myr). Future numerical simulations following the evolution of disks in less dense regions are needed for a more accurate quantitative comparison between predictions from the ejection scenario and the results of the observations presented here.


It is important to stress that the sample of BDs and VLMs investigated so far at high angular resolution, other than being limited to a handful of objects only, is also strongly biased toward the brightest disks which can be spatially resolved more easily at sub-mm wavelengths. The fact that none of these disks has revealed a very small outer disk radius ($R_{\rm{out}} < 20$ AU) as predicted by ejection models may just reflect this selection bias, as the large population of fainter disks may have lower fluxes because of smaller disk radii. 
Future ALMA observations of VLMs/BDs will characterize disk sizes for large samples of disks with sub-mm fluxes comparable and lower than those presented here and will test the viability of the ejection scenario to form very low mass stars and brown dwarfs.

\subsection{The potential of VLMs/BDs disks to form rocky planets and gas giants}

\noindent Understanding how dust and gas are distributed across the disk is important also for models of planet formation, as the physical structure of the disk determines the efficiency of planet formation at different semi-major axes as well as of radial migration. 
This is particularly critical for disks surrounding very low mass stars and brown dwarfs, for which their low mass limits the mass of the largest planets that can be assembled. 

\citet{Payne:2007} investigated the potential of planet formation in disks around brown dwarfs by extending to lower masses the calculations by \citet{Ida:2004,Ida:2005} in the framework of the sequential core accretion model.
These models start with a distribution of planetesimals embedded in a gas rich disk. The core accretion of planetesimals on the forming protoplanet is modeled using the combined results of analytic models and $N$-body simulations. Once the cores reach a critical mass, the accretion of a gas envelope starts and is regulated by Kelvin-Helmholtz contraction. The models take into account also the effects of radial migration. The authors use this framework to explore the effect on the planetary mass-semimajor axis distribution of varying physical parameters, such as the surface density radial profile and total mass of the disk.

They find that for disks with a total mass of a few $M_{\rm{Jup}}$ surrounding a 0.05~$M_{\odot}$ brown dwarf, giant planets cannot be formed, in the sense that no rocky cores in their simulation accrete a significant gaseous envelope. The three disks considered in this paper have estimated masses of $\sim 1 - 2~M_{\rm{Jup}}$ (Table~\ref{tab:disk_properties}) and are among the most massive disks around objects with masses $\sim 0.05 - 0.1~M_{\odot}$ observed so far at mm-wavelengths \citep[][]{Bouy:2008,Klein:2003,Scholz:2006,Mohanty:2013,Ricci:2013,Andrews:2013}. This suggests that the core accretion mechanism should be very inefficient at forming giant planets around brown dwarfs and very low mass stars. Any giant planetary mass companion found around these very low mass objects are likely formed via other mechanisms \citep[see e.g. the case of 2M1207b,][]{Chauvin:2005, Lodato:2005}.

At the same time, \citet{Payne:2007} show that brown dwarf disks with masses of few $M_{\rm{Jup}}$ have the potential to form rocky planets, and the mass of the most massive planet which can be formed in these disks critically depends on the slope $p$ of the radial distribution of dust. At a given mass, disks with steeper (decreasing) radial profiles can produce more massive planets as the surface densities in the inner disk where planets form are higher than for disks with shallower radial density profiles.  
They find that for $p = 1.5$ rocky planets with masses of $\approx 1 - 5~M_{\oplus}$ can be found at distances of $\approx 0.2 - 3$ AU from the brown dwarf. For a lower value of $p = 1.0$, the mass of the planets that can be formed at these orbital radii is decreased by a factor of 5 to 10.  

Under the assumption that the $p$-values constrained for our disks extend toward the inner disk regions, i.e. within the spatial resolution of our ALMA observations ($\approx 30$ AU in radius), the relatively large values of $p$ derived by our analysis ($\approx 1.20 - 1.65$, see Table~\ref{tab:disk_properties}) together with the relatively large masses of the disks indicate that rocky planets with masses $> 1~M_{\oplus}$ can be formed in these disks.
This is consistent with the detection of a $\sim 3~M_{\oplus}$ super-Earth planet orbiting at a distance of 0.7 AU from the very low mass star MOA-2007-BLG-192L \citep[$M_{\star} = 0.084^{+0.015}_{-0.012}~M_{\odot}$,][]{Bennett:2008,Kubas:2012}.

\section{Summary}

\noindent We presented continuum and spectral line data for three disks surrounding young brown dwarfs and very low mass stars (2M0444, CIDA 1, CFHT Tau 4) in the Taurus star forming region using ALMA in Early Science (Cycle 0) at 0.89~mm and 3.2~mm. Dust thermal emission is detected at both wavelengths and spatially resolved at 0.89 mm for all the three disks, while CO($J=3-2$) emission is seen in two disks, 2M0444 and CIDA 1. The velocity maps of the CO($J=3-2$) emission suggest molecular gas in rotation around the central objects. 
 
We analyzed the continuum interferometric visibilities at 0.89 mm using physical models of disks and constrain the disks physical structure in dust. The results of our analysis show that the disks are relatively large, with outer radii of $\approx 140$ AU, 70 AU and a lower limit of 80 AU for 2M0444, CIDA 1 and CFHT Tau 4, respectively. The inferred disk radii, radial profiles of the dust surface density ($\Sigma \propto r^{-p}$, with $p \approx 1.2 - 1.7$) and disk to central object mass ratios ($\approx 0.8 - 2.6\%$) lie within the ranges found for disks around young stars. 

These results suggest a stellar-like gravitational collapse of a dense pre-(sub)stellar core as the the most plausible  mechanism to form these three very low mass objects. Numerical simulations for the formation of brown dwarfs and very low mass stars via dynamical ejections in dense stellar clusters tend to create preferentially smaller disks than those presented in this paper \citep[][]{Bate:2003,Bate:2012}. However, future simulations following the disk evolution to $\sim$ Myr ages in lower density stellar associations are necessary for a more accurate comparison with our observational results within this theoretical framework. 

The spectral profile of the disk sub-mm/mm SED, together with the large radii inferred from the interferometric visibilities, indicate that grains with at least millimeter sizes are present in these disks. This result confirms the recent finding that the evolution of solids up to these sizes proceeds in a similar way in disks around brown dwarfs/very low mass stars as in disks around young stars \citep{Ricci:2012a}. A complete understanding of the size evolution and dynamics of these small solids in protoplanetary disks is still beyond our reach, and our results favor assumptions/mechanisms that boost the efficiency of dust coagulation even in low-density environments \citep[e.g. relatively large fragmentation velocities, low turbulent viscosities, distribution functions for the grain velocities, mass transfer through collisions between grains with very different mass;][]{Pinilla:2013,Meru:2013}, as well as mechanisms which can efficiently slow down the otherwise fast inner migration of these grains~\citep[e.g. local pressure bumps,][]{Pinilla:2013}. 
Since these pebbles are expected to play a key role in the formation of planetesimals in gas-rich disks \citep[for a recent review, see][]{Johansen:2014}, our results also suggest that debris disks around brown dwarfs should be common.
 
Finally, according to the \citet{Payne:2007} models of planet formation within the core accretion scenario, disks with the properties of the objects presented here should not form giant planets but are capable of forming rocky planets with masses up to a few $M_{\oplus}$ around brown dwarfs and very low mass stars \citep{Bennett:2008, Kubas:2012}. Given that our observations probe the high-mass tail of the known disks around these objects, this suggests that the core accretion mechanism should be very inefficient at forming giant planets around these very low mass objects. Any giant planetary mass companion found around them is likely formed via other mechanisms~\citep[][]{Lodato:2005}. 
Future observations with ALMA will allow to test some of the sensitive assumptions in this discussion, e.g. the ISM-like gas to dust mass ratio in the disk.

\acknowledgments We thank E. van Kampen and ESO ARC for technical support.
I. G. is supported by the Spanish MINECO grant AYA2011-30228-C03-01 (co-funded with FEDER fund). 
This paper makes use of the following ALMA data: ADS/JAO.ALMA\#2011.0.00259.S. ALMA is a partnership of ESO (representing its member states), NSF (USA) and NINS (Japan), together with NRC (Canada) and NSC and ASIAA (Taiwan), in cooperation with the Republic of Chile. The Joint ALMA Observatory is operated by ESO, AUI/NRAO and NAOJ. The National Radio Astronomy Observatory is a facility of the National Science Foundation operated under cooperative agreement by Associated Universities, Inc.


\end{document}